\documentclass[preprint,journal]{vgtc}       

\ifpdf
  \pdfoutput=1\relax                   
  \usepackage{graphicx}                
  \DeclareGraphicsExtensions{.pdf,.png,.jpg,.jpeg} 
\else
  \usepackage{graphicx}                
  \DeclareGraphicsExtensions{.eps}     
\fi%

\graphicspath{{figures/}{pictures/}{images/}{./}} 

\usepackage{microtype}                 
\PassOptionsToPackage{warn}{textcomp}  
\usepackage{textcomp}                  
\usepackage{mathptmx}                  
\usepackage{times}                     
\usepackage{cite}                      
\usepackage{tabu}                      
\usepackage{booktabs}                  

\usepackage{marginnote}
\usepackage{stmaryrd}

\usepackage{mathtools, nccmath, amssymb}

\usepackage[ruled,linesnumbered]{algorithm2e} 

\usepackage[textsize=scriptsize]{todonotes} 
\usepackage[autostyle=true,german=quotes]{csquotes} 

\usepackage{subfigure}

\usepackage{multirow}
\usepackage{hhline}

\usepackage[final]{changes}

\makeatletter
\g@addto@macro\normalsize{%
	\setlength\abovedisplayskip{5pt}
	\setlength\belowdisplayskip{5pt}
	\setlength\abovedisplayshortskip{5pt}
	\setlength\belowdisplayshortskip{5pt}
}
\makeatother

\ieeedoi{10.1109/TVCG.2020.3030436}

\onlineid{0}

\vgtccategory{Research}
\vgtcpapertype{algorithm/technique}

\title{Homomorphic-Encrypted Volume Rendering}  

\author{Sebastian Mazza, Daniel Patel, and Ivan Viola}
\authorfooter{
\item
 Sebastian Mazza is with TU Wien, Austria. E-mail: sebastian@mazza.at.
\item
 Daniel Patel is with Western Norway University of Applied Sciences, Norway. E-mail: danielpatel.no@gmail.com.
\item
 Ivan Viola is with King Abdullah University of Science and Technology (KAUST), Saudi Arabia. E-mail: ivan.viola@kaust.edu.sa.
}

\shortauthortitle{Mazza et al.: Homomorphic-Encrypted Volume Rendering}

\abstract{Computationally demanding tasks are typically calculated in dedicated data centers, and real-time visualizations also follow this trend.
Some rendering tasks, however, require the highest level of confidentiality so that no other party, besides the owner, can read or see the sensitive
data.
Here we present a direct volume rendering approach that performs volume rendering directly on encrypted volume data by using the homomorphic Paillier encryption algorithm.
This approach ensures that the volume data and rendered image are uninterpretable to the rendering server.
Our volume rendering pipeline introduces novel approaches for encrypted-data compositing, interpolation, and opacity modulation, as well as simple transfer function design, where each of these routines maintains the highest level of privacy.
We present performance and memory overhead analysis that is associated with our privacy-preserving scheme.
Our approach is open and secure by design, as opposed to secure through obscurity.
Owners of the data only have to keep their secure key confidential to guarantee the privacy of their volume data and the rendered images.
Our work is, to our knowledge, the first privacy-preserving remote volume-rendering approach that does not require that any server involved be trustworthy; even in cases when the server is compromised, no sensitive data will be leaked to a foreign party.
} 


\CCScatlist{ 
 \CCScat{K.6.1}{Management of Computing and Information Systems}%
{Project and People Management}{Life Cycle};
 \CCScat{K.7.m}{The Computing Profession}{Miscellaneous}{Ethics}
}

\teaser{
  \centering
  \includegraphics[width=\linewidth,trim=0 2mm 0 3mm,clip]{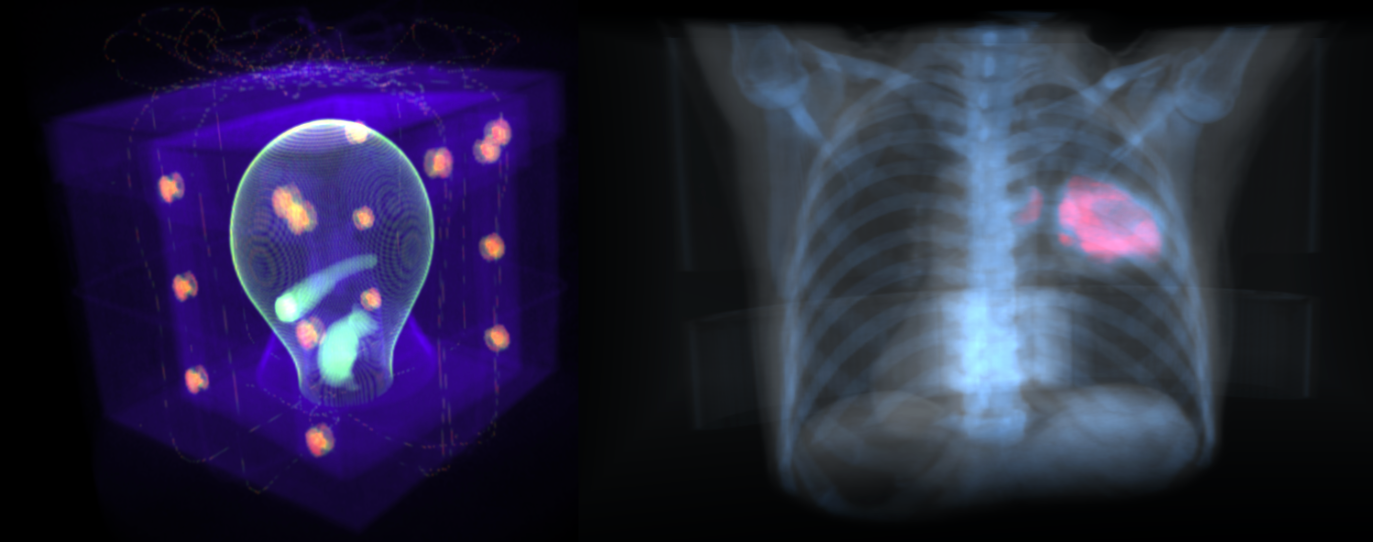}
  \caption{
		\replaced{
			Two decrypted images rendered by our simplified transfer function approach.
			The left image is rendered from a CT scan of a water globe and other objects inside a box that is wrapped inside a present box \cite{datasetPresent}.
			The dataset used for rendering the right image is a CT/PET scan of the thorax from a patient with lung cancer
			({\em G0061/Adult-47358/7.0} from the {\em Lung-PET-CT-Dx} dataset  \cite{datasetLungCancer, article:datasetTCIA}).
		}{
			Two decrypted images rendered by our simplified transfer function approach from the same encrypted dataset.
			The dataset is a CT scan of water globe and other objects inside a box that is wrapped in a present box \mbox{\cite{datasetPresent}}.
		}
	}
  \label{fig:teaser}
	\vspace{-0.2cm}
}

\vgtcinsertpkg

\SetCommentSty{small}

\begin{document}


\firstsection{Introduction}

\maketitle

Volume rendering is extensively used in domains where the underlying data is considered highly confidential. One  example includes the field of medicine, where CT, MRI, or PET data are used for diagnostic or treatment-planning purposes.
Another such example is hydrocarbon and mineral exploration in energy industries for inspecting the subsurface using seismic scans.

For volume rendering, privacy can currently only be achieved by storing and processing the datasets locally.
Volume rendering requires computers with large memory and powerful processing power.
Such hardware must be frequently maintained and upgraded.
Therefore, for many organizations, it would be advantageous to outsource the rendering to cloud services.
As cloud services remove the need to be in close proximity to the rendering hardware, users can now also view volume rendering on thin clients that do not have the required memory or processing power, such as tablets and smart phones.
However, hospitals must protect sensitive personal data and energy companies must protect their valuable data assets.
Thus, it is essential that their data is not visible to the cloud services, as these either cannot be trusted, or their security might be compromised.
\added{
	Therefore, we want to make it possible to perform direct volume rendering on untrusted hardware while preserving the same level of privacy for the datasets as the privacy achieved with a classical local rendering approach.
}

\begin{figure*}[t]
	\centering
	\includegraphics[width=\textwidth]{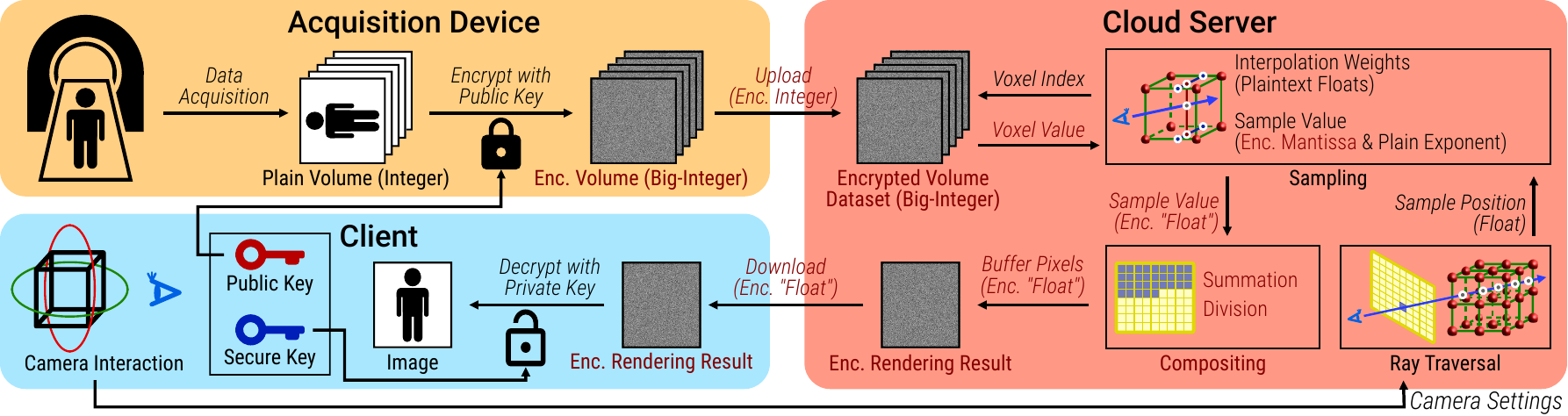}
	\caption{
	Our approach consists of a computer that produces, encrypts, and sends volume data to a server, which then renders the data and sends the result to a client. The client decrypts and visualizes the result.
	\added{The text that belongs to encrypted data or processing is stated in red.}
	}
	\label{fig:setup}
\end{figure*}

The basic concept of our privacy-preserving approach is shown in \autoref{fig:setup}.
First, the data is acquired and immediately encrypted by, for example, a machine that is directly connected to a medical scanner.
Then the encrypted volume is uploaded to the {\em honest-but-curious}\footnote{The server will perform the algorithm it is requested to compute honestly (correctly); however,  there is the potential for possible access by the curious eyes of administrators or hackers.}\cite{honestButCurious} public server.
This is done only once per volume.
When the clients that hold the secure key request rendering, the server performs ray-casting directly on the encrypted volume data.
This computation results in an image containing encrypted values, which is then sent to the client.
When the client receives the requested image, it is decrypted and displayed to the user.
As the server that computed the rendered image will only see encrypted data, our approach maintains privacy.

\replaced{
	Our design is constrained by three requirements.
	The {\em first requirement} is that the privacy of the user data is protected by the design of the algorithm
	and does not depend on hiding implementation details, keeping any part of the system secret, or any other obscure technique
	that cannot be secure, at least not in the long run (Kerckhoffs's principle \cite{article:Kerckhoffs83}).
	Such obscure techniques only make it difficult to know the actual security of the system.
	Therefore, the security of our volume rendering approach solely depends on the security of a well-established cryptographic algorithm, continuously being scrutinized in cryptographic research.
	We have chosen the well-established Paillier cryptographic algorithm, which is partly homomorphic \cite{pdf:Paillier}.
}{
	Our objective is to ensure that the security of our volume rendering approach solely depends on the security of a well-established cryptographic algorithm,  continuously being scrutinized in cryptographic research.
	The security of our approach does not rely on hiding details, as this only makes it hard to know the actual security of the system.
	We have chosen the well-established Paillier cryptographic algorithm, which has the added benefit of being partly homomorphic.
}
The key property of homomorphic encryption (HE) is that
arithmetic operations on encrypted data are dual to arithmetic operations on
\emph{plaintext} (original, unencrypted)
data.
This enables an algorithm to perform a correct 3D volume rendering image synthesis directly on the encrypted data, without being able to ever access the plaintext data.
\replaced{
	As a consequence, the result of the rendering on the server is an encrypted image.
}{
	The server produces resulting renderings as encrypted images.
}

\added{
We are currently not able to show interactive frame rates with
the proof-of-concept implementation of our approach.
However, one future research goal should attempt to make a remote rendering system fast enough to achieve this.
This leads to our {\em second requirement}, which is to only use techniques that will not prevent the system from scaling the performance with the computational power available on the server and will not prohibit interactive frame rates.
The {\em third requirement} is to support thin clients without much memory and computational power.
As a result, we consider the client to be a low powered device, which is connected to a mobile or another medium-bandwidth network, while we assume that the server is a powerful machine (e.g., with multiple professional GPUs) or even a compute cluster.
}

\added{
	By using encryption schemes like AES \cite{NIST:AES}, it is currently possible to store volume datasets securely in the cloud.
	However, for rendering images from the datasets, the entire volume needs to be downloaded and decrypted first, and then rendered on the client.
	A privacy-preserving remote volume rendering can also make the cloud more attractive as a storage space for volume data, because with our proposed technique, it is no longer necessary to download the whole dataset before images can be synthesized from it.
}

\section{Related Work}
We have only found two  works that address the topic of privacy-preserving rendering of volumetric data.
The most similar work to ours is that of Mohanty et al.\cite{inproc:3DCrypt}.
They present a cryptosystem for privacy-preserving volume rendering in the cloud.
Unlike our approach, they achieve correct alpha compositing. 
However, to attain this goal, they end up with a solution that cannot be considered secure, that has a fixed transfer function, and that requires that the volume is sent from one server to another server for each rendered frame.

Their approach requires two servers for rendering: a Public Cloud Server and a Private Cloud Server.
The first step of their rendering approach is to apply a color and opacity to each voxel before encrypting the volume. 
This means that the transfer function is pre-calculated and cannot be changed by a user without performing a time-consuming reencryption and uploading of the volume.
In the next step, the encrypted data is uploaded to the Public Cloud Server, which stores the volume data.
When the Public Cloud Server receives an authorized rendering request from a client, the server calculates all sample positions for the requested ray casting and interpolates the encrypted color and opacity values for each sample position.
All interpolated sample values then need to be individually sent to the Private Cloud Server, which decrypts the opacity value of each sample in order to perform the alpha blending along the viewing rays.
For alpha compositing, the opacity values of samples represent object structures in the volume; therefore, anyone who can gain access to the Private Cloud Server, such as an administrator or a hacker, will be able to observe these structures in the volume dataset.
If an unauthorized person has access to this server,
the whole approach collapses.
For the task of encrypting and decrypting parts of the volume data on the servers, their approach requires a central Key Management Authority (KMA).
While this brings the advantage that an organization can centrally control which users have access to a specific volume, it enlarges the attack surface of their system considerably, because the KMA has all keys required for decrypting all volume data.
Therefore, the confidentiality of the KMA is constitutional for the privacy of all datasets, no matter who they belong to.

Another weakness of their approach is the required network bandwidth between the Public and the Private Cloud Server because all sample values of a ray casting frame need to be transferred from the Public and the Private Cloud Server (more than 1GB).
With our approach, the privacy of the volume data and rendered image depends only on a single secure key. Also, our approach should scale linearly with the computing power of the hardware it is running on.

Chou and Yang \cite{ObfuscatedVolRend} present a volume rendering approach that attempts to make it difficult for an unintended observer to make sense of the volume dataset that resides on a server.
This is done by, on the client's side, subdividing the original data into equally sized blocks.
The blocks are rearranged in a random order and then sent to the server as a volume.
The server then performs volume rendering on each block and sends the result back to the client, which will reorder the individual block renderings and composite them to create a correct rendering.
To obfuscate the data further, on the client's side, the data values in each block are changed using one out of three possible monotonic operations: flipping, scaling, and translating.
Monotonic operations are used as they are invertible and associative under the volume rendering integration.
Therefore, doing the inverse operators on the resulting rendering gives the same result as doing them on the data values before performing the rendering.
This algorithm cannot be considered safe, and the authors acknowledge this as they state that the goal is only to not trivially reveal the volume to unauthorized viewers.
A possible attack would be to consider the gradient magnitude of the obfuscated volume.
This should reveal the block borders.
The gradient magnitude can further be used inside each block to reveal structures in the data that can be used for aligning the blocks correctly.

\deleted{
Our goal is to develop an approach that is open and secure by design
(Kerckhoffs's principle \cite{article:Kerckhoffs83}) and not {\em secure through obscurity} \cite{article:hoepman2008securityThrough}
or relies on assumptions that are outside the algorithm itself, such that the internal memory of a cloud server is not accessible to an intruder.
The former is the case for Chou and Yang's approach \cite{ObfuscatedVolRend} and the latter is the case for the approach by Mohanty et al. \cite{inproc:3DCrypt}.
}

\replaced{
	To attain our goal of developing an approach that is open and secure by design,
	we use the Paillier cryptosystem developed by Paillier in 1999 \cite{pdf:Paillier}.
	This cryptosystem  is an
	asymmetric encryption scheme,
	where the secure key contains two large prime numbers $p$ and $q$%
	, and the public key contains the product $N$ (modulus)
	of $p$ and $q$.
	The cryptosystem supports an additive homomorphic operation ($\oplus$).
	If this operation is applied to two encrypted values $\llbracket m_1 \rrbracket, \llbracket m_2 \rrbracket$ ($\llbracket m \rrbracket$ means encrypted $m$), the decrypted result is the sum of the $m_1$ and $m_2$ ($\textrm{Dec}(\llbracket m_1 \rrbracket \oplus \llbracket m_2 \rrbracket) = (m_1 + m_2) \mod N$).
	Furthermore, a homomorphic multiplication ($\otimes$) between an encrypted value and a plaintext value $d$ is supported
	($\textrm{Dec}(\llbracket m_1 \rrbracket \otimes d) = (m_1 \times d) \mod N$).
	Since Paillier's cryptosystem does not carry over  multiplication of two encrypted values to plaintext, it is classified as a partially homomorphic encryption (PHE) scheme.
	Paillier can securely encrypt many values (e.g., $512^3$ voxels of a volume) from a small number space (e.g., $2^{10}$ possible density values), because it is {\em probabilistic}, which means that during the encryption, the {\em obfuscation} can map a single plaintext value randomly to a large number of possible encrypted values. 
	This makes a simple \enquote{probing} for finding out the number correspondence
	impossible.
	Further details about Paillier's cryptosystem such as the encryption and decryption algorithm is provided in the Supplementary material document.
	We are limited to the arithmetic operations supported by Paillier for creating a volume rendering that captures as much structure as possible from the data.
} {
To attain this goal, we use the Paillier cryptosystem, and are limited in the types of arithmetic operations we can use for creating a volume rendering that captures as much structure as possible from the data.
}
This forces us to think unconventionally and creatively when designing the volume renderer.

For homomorphic image processing, the work by Ziad et al. \cite{pdf:CryptoImg-IEEE} makes use of the additive homomorphic property of Paillier's cryptosystem. They demonstrate that they are able to implement many image processing filters using the limited operations allowed with Paillier. They implement filters for negation, brightness adjustment, low pass filtering, Sobel filter, sharpening, erosion, dilation and equalization.
\added{
While most of these filters are computed entirely on the server side, erosion, dilation, and equalization require the client for parts of the computation.
There are various works that make use of such a {\em trusted client protocol} approach to overcome the limitation of a
PHE
scheme
and enable operations such as addition, multiplication, and comparisons on the encrypted data~\cite{inProc:SecureMR, inProc:Crypsis, inProc:JCrypt}.
A {\em trusted client} knows the secure key and can, therefore, perform any computation on the data or {\em convert / re-encrypt} it from one encryption scheme to another (e.g., from an additive to a multiplicative {\em homomorphic encryption}).
These client-side computations introduce latency because the data needs to be transferred back and forth between the server and the client.
Furthermore, the client needs to have enough computational power to avoid becoming the bottleneck of the system.
To mitigate this problem, automated code conversions can be used that minimize the required client side {\em re-encryptions} \cite{inProc:JCrypt,inProc:SecureMR}.
While a {\em trusted client} approach could theoretically solve many of the problems we face with our untrusted server-only approach, it is not practical for volume rendering.
The most demanding problems of volume rendering, such as transferring a voxel value and advanced compositing (alpha blending, maximum intensity projection, ...), need to be done per voxel.
Hence, every voxel that could contribute to the image synthesis (all voxels of a volume for many rendering cases) needs to be transferred to the {\em trusted client} and processed there for every rendered frame.
The encryption and decryption on the client side are more expensive than the operations required for a classical sample compositing due to the size of encrypted values (e.g., 1000 bit per voxel).
If an amount of data in the range of the volume itself needs to be transferred from the server to the client, where the data would need to be encrypted and decrypted, it is pointless to perform any calculations on the server, because the client then has more work to do than in a classical volume rendering on the client.
Moreover, it does not save any network traffic as compared to a simple download, decrypt, and process use case.
Therefore, we argue that {\em trusted client} approaches are not suitable for our work.
Furthermore, a {\em trusted client} approach will not work with thin clients, which contradicts our third requirement.
Our second requirement is also contradicted because, in real-world use cases, the network bandwidth between a client like a tablet computer and a cloud server will not have enough bandwidth (e.g., more than 1Gbit/s) to support interactive frame rates.
}


\section{Encrypted Rendering Overview}
The first step of the introduced privacy preserving rendering system is the encryption of the volume dataset \added{(\autoref{fig:setup} Acquisition Device)}.
During the encryption stage, every single scalar voxel value of a volume dataset needs to be encrypted with Paillier's approach (see \autoref{alg:PaillierEncrypt} in the Supplementary Material document).
Meta data of the volume such as width, height, depth and the storage order of voxels will not be encrypted.
The next step is to upload the encrypted volume dataset to a server \added{(\autoref{fig:setup} arrow from Acquisition Device to Cloud Server)}.
For our approach, the device that encrypts the volume and uploads it to a server does not even need the secure key, because for encryption, only the public key is required.

When a rendered image is requested to be shown on a client, the client sends a rendering request to the server, which has the encrypted volume dataset \added{(\autoref{fig:setup} arrow from Client to Cloud Server)}.
The rendering request contains further information about the settings of the rendering pipeline, such as the camera position, view projection, and (depending on the selected rendering type) also information about the transfer function that should be used.
After the server receives such a rendering request, it uses the included pipeline settings and the already stored encrypted volume dataset to render the requested image \added{(\autoref{fig:setup} the rendering pipeline stages of the Cloud Server)}.
To preserve privacy, the server does not have the secure key and can not, therefore, decrypt the volume data.
The operations that are used for rendering an image from an encrypted volume dataset are limited to the homomorphic operations add ($\oplus$\deleted{, \autoref{equ:paillierAddition}}) and multiply with plaintext ($\otimes$\deleted{, \autoref{equ:paillierMultiplication}}), which are defined for Paillier's encryption scheme.
When the rendering is finished, the server will send the calculated image data to the client \added{(\autoref{fig:setup} arrow from Cloud Server to Client)}.
The resulting image that the client receives is still encrypted.
Decrypting such an image is only possible for a client that knows the correct secure key.
For everyone else, the image will be random noise (shown in Supplementary Video Material).
Since every single pixel value is an encrypted number, every single pixel can be decrypted independently of the other pixels.
For a gray-scale image, that means one number per pixel.
An RGB colored image requires three values that need to be decrypted per pixel.

In \autoref{sec:xray}, we explain how the homomorphic operations of Paillier's HE can be used for X-ray sample integration. Furthermore, we will show how to use Paillier's cryptosystem with floating-point numbers, which allows us to perform trilinear interpolation.
\autoref{sec:transferFunction} explains a more advanced approach that allows the emphasizing of different density ranges in the rendered images.

\section{Encrypted X-Ray Rendering}
\label{sec:xray}
\added{
	Ray-casting \cite{proc:Krueger:2003:ATGV} is the most frequently used approach for volume rendering.
	Furthermore, ray-casting based algorithms can be easily and efficiently parallelized and can be implemented with fewer memory reads than slicing-based algorithms.
	Memory access is time-consuming, especially if every number that needs to be read is thousands of bits long.
	Therefore, we implement our privacy-preserving volume rendering approach with ray-casting.
}
\replaced{
	However, other direct volume rendering approaches developed for unencrypted data, such as slicing, can be used as well.
	Slicing on the server can be built by the same encrypted rendering pipeline components (sampling / interpolation, color mapping, compositing), 
	which we will explain anon.
	Slicing could also be used to just perform the sampling on the server, transfer the slices to the client, and perform the compositing there.
	However, this would not fulfill our
	requirements because of the required network bandwith and the high computational requirement on the client.
}{
}

\added{
The ray casting algorithm first calculates a viewing ray for every pixel of the final image (\autoref{fig:setup} Ray Traversal - stage of the Server).
These viewing rays will be calculated based on the camera position, up vector, opening angle, image resolution, and pixel index. 
At discrete and equidistant steps along the ray, the data of the volume is sampled (\autoref{fig:setup} Sampling - stage of the Server).
The last step is the compositing, where the final pixel value is calculated based on the sample values of a viewing ray (\autoref{fig:setup} Compositing - stage of the Server).
}

\replaced{
	X-ray rendering is a volume rendering approach where the sample value is mapped to a white color with monotonically increasing opacity, and the compositing is a summation followed by a normalization at the end of the ray traversal.
}{
}
If the sampling of the voxel values is done by nearest-neighbor filtering, the sum along a viewing ray can be calculated by only using the homomorphic add operation ($\oplus$) which is already defined for Paillier's cryptosystem\deleted{ (see \autoref{equ:paillierAddition})}.
\replaced{
The final normalization of all samples along a view ray cannot be done directly by the homomorphic operations of Paillier's encryption scheme because this requires a division that can result in a non-integer value that is not supported.
}{
}
However, the server
\replaced{could}{}
send the encrypted sum together with the sample count to the client, which can perform the division after decrypting the sum.

\begin{figure}[tb]
	\centering
	\subfigure[Nearest Neighbor]{
		\includegraphics[width=0.478\columnwidth,trim=0 3mm 0 3mm,clip]{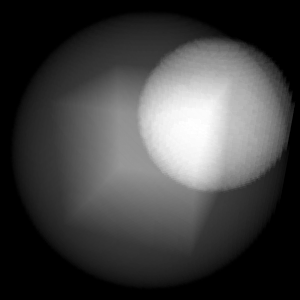}
		\label{fig:xrayTrilinear:nn}
	}
	\subfigure[Trilinear Interpolation]{
		\includegraphics[width=0.478\columnwidth,trim=0 3mm 0 3mm,clip]{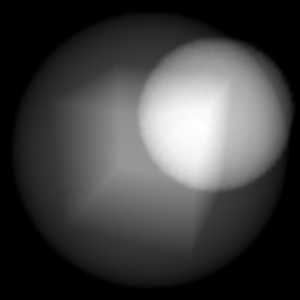}
		\label{fig:xrayTrilinear:ti}
	}
	\caption{Results from encrypted X-ray rendering showing nearest neighbor (a) and Trilinear interpolation (b), which we also support.}
	\label{fig:xrayTrilinear}
\end{figure}

\added{
To improve the nearest-neighbor sampling with trilinear interpolation,
	a mechanism that allows the summing and normalization of encrypted values ($\llbracket m_1 \rrbracket$, $\llbracket m_1 \rrbracket$), which are scaled by some plaintext weights ($\alpha_1$, $\alpha_2$), is required.
	For plaintext integers, the interpolation could be implemented
	around the integer arithmetic operations add, multiply, and divide
	(1D example: $(m_1 \cdot \alpha_1 + m_2 \cdot \alpha_2) / (\alpha_1 + \alpha_2)$).
	Since an arbitrary division is not supported by Paillier's cryptosystem, this is not directly feasible on encrypted data.
	A possible solution could be to use fraction types, which has an encrypted denominator and a plaintext numerator for storage and calculations.
	After the image is rendered, which contains such fractions as pixel values,
	 the client can download it, decrypt the denominators, and perform the deferred divisions%
}%
\footnote{
	\added{
		If the rendering pipeline is designed in a very static way, it is theoretically possible to know the final numerator upfront and let the client perform the required division without explicitly specifying the numerator. However, this is very inflexible, error prone, and requires an update for the client whenever a change on the server leads to a change of the final numerator.
	}
}.
\added{
	However, we decided to use a floating-point encoding, which is easier to implement
	and allows a shader code development as is usual for hardware accelerated rendering.
	With a floating-point representation of encrypted values, it is possible to multiply the eight neighboring voxels of a sample position with the distances between the samples and voxel position. These distances, which have a sum of $1.0$, are the weights of the interpolation (1D example: $m_1 \cdot \alpha_1 + m_2 \cdot \alpha_2$).
}
\deleted{
To improve the nearest-neighbor sampling with trilinear interpolation, a floating-point number encoding for encrypted values is required because the distances between the sample position and the eight neighboring voxels, which are used as weights for the interpolation, are fractions.
}
A floating-point encoding will also make the final division of the sample sum for X-ray rendering on the server side possible.
While a floating-point encoding does not directly enable divisions in the encrypted domain, it can be used to approximate a division by a multiplication with the reciprocal of the divisor, as shown in \mbox{\autoref{equ:fpDeviation}}.
\begin{equation}
	\begin{split}
	\frac{\sum}{n}
	\approx \textrm{Dec}\left(
				\left\llbracket\sum\right\rrbracket
				\otimes
				\left\lfloor \frac{1}{n} \cdot 10^\gamma \right\rceil
			\right)
		\cdot 10^{-\gamma}
	\end{split}
	\label{equ:fpDeviation}
\end{equation}
The sum of samples along a viewing ray is denoted as $\sum$, and $n$ is the count of samples.
The precision of the approximation is defined by the count of decimal digits $\gamma$ (e.g., $\gamma = 3$ for thousandth).
Before the reciprocal of $n$ is multiplied with $\sum$, the comma is moved $\gamma$ digits to the right ($\cdot 10^\gamma$) and then rounded ($\lfloor \rceil$).
The multiplication with $10^{-\gamma}$, which moves the comma back to the correct position, can be achieved by subtracting $\gamma$ from the exponent of the floating-point encoded result.
Since the Paillier cryptosystem is defined over $\mathbb{Z}_N$, the result is only correct if no intermediate result is greater than $N-1$.

We will discuss the used floating-point encoding in \autoref{sec:FPNumbers}.
\autoref{fig:xrayTrilinear} shows two images that were rendered from an encrypted floating-point encoded dataset.
For the rendering of the left image, a nearest-neighbor sampling was used, and for the right image, a trilinear interpolation was used.
The used dataset contains three objects with different densities: a solid cube in the center wrapped inside a sphere and another sphere at the top left front corner.
The same dataset is also used for renderings shown in \autoref{fig:densityEmphasising} and \autoref{fig:generic4dimColored}.

\subsection{Encrypted Floating-Point Numbers}
\label{sec:FPNumbers}
A floating-point number is defined as $m \cdot b^e$, where
$m$ is called the mantissa.
The exponent $e$ defines the position of the comma in the final number.
The base $b$ is a constant that is defined upfront (e.g., during the compilation of the application).
We used a decimal system for convenience; therefore, our prototype uses $b = 10$.
However, $b$ can be any positive integer that is greater or equal to $2$.

To calculate with floating-point arithmetic in the encrypted domain, we have chosen to use the approach developed for Google’s Encrypted BigQuery Client \cite{web:GoogleEncryptedBigqueryClientGit}.
The idea is to store the mantissa $m$ and the exponent $e$ of a floating-point number in two different integer variables.
During the encryption of the floating-point number $(m, e)$, only the mantissa $m$ is encrypted using Paillier's cryptosystem.
The exponent $e$ remains unencrypted, which results in the floating-point number $(\llbracket m \rrbracket, e)$.
This floating-point number representation is also used by the {\em python-paillier} library \cite{web:PythonPaillierGit}, the Java library {\em javallier} \cite{web:PythonPaillierGit} and in the work by Ziad et al. \cite{pdf:CryptoImg-IEEE}.

For an addition of two such encrypted floating-point numbers, both need to have the same exponent.
Therefore, the exponents of both numbers must be made equal before the actual addition, if they are not already equal.
Hence, it is not possible to increase the exponent if the mantissa is encrypted because that would require a homomorphic division of the encrypted mantissa, which is not possible.
Therefore, the floating-point number with the greater exponent needs to be changed.
On the other hand, decreasing the exponent of a floating-point number is not a problem because it requires a homomorphic multiplication of the encrypted mantissa with a plaintext number, which is possible with Paillier.
\autoref{equ:decreaseExponentTo} shows how to calculate the new mantissa $\llbracket m_n \rrbracket$ that is required for decreasing the exponent of the floating-point number ($\llbracket m_o \rrbracket$, $e_o$) to the lower exponent $e_n$.
The new floating-point number is defined as ($\llbracket m_n \rrbracket$, $e_n$), which represents exactly the same number as ($\llbracket m_o \rrbracket$, $e_o$).
It is just another way to store it.
\begin{align}
\begin{split}
	\llbracket m_n \rrbracket = \llbracket m_o \rrbracket \otimes b^{e_o - e_n}
\end{split}
\label{equ:decreaseExponentTo}
\end{align}
When both floating-point numbers ($\llbracket m_1 \rrbracket$, $e_1$) and ($\llbracket m_2 \rrbracket$, $e_2$) have the same exponent $e_1 = e_2 = e_n$, the homomorphic sum $\llbracket m_s \rrbracket$ of both mantissas can be calculated by the add operation defined for Paillier \deleted{(see \autoref{equ:paillierAddition}) }, which results in the final floating-point number ($\llbracket m_s \rrbracket$, $e_n$).
The \autoref{alg:PaillierFpAdd} shows this approach for summing two floating-point numbers with encrypted mantissas.
The lines from 2 to 10 bring the exponents of both floating-point numbers to the same value ($e_n$), and line number 11 contains the addition of the encrypted mantissas.
%

\begin{algorithm}[tb]
\DontPrintSemicolon

\SetKwInOut{Param}{Parameters}
\SetKw{TypeInt}{integer}
\SetKwProg{Proc}{procedure}{}{}

\SetKwFunction{FnPfAdd}{fpAdd}

\BlankLine
\Param{Encrypted mantissas $\llbracket m_1 \rrbracket, \llbracket m_2 \rrbracket$ and plaintext exponents $e_1, e_2$ of the two floating point numbers that should be summed. \\ $b$ is the used base, e.g. 10 for a decimal system.}
\KwResult{Encrypt mantissa $\llbracket m_s \rrbracket$ and plaintext exponent $e_n$.}
\BlankLine

\Proc{\FnPfAdd{$\llbracket m_1 \rrbracket$, $e_1$, $\llbracket m_2 \rrbracket$, $e_2$}}{
	\uIf{$e_1 > e_2$} {
		$\llbracket m_1 \rrbracket = \llbracket m_1 \rrbracket \otimes b^{e_1 - e_2}$ \;
		$e_n = e_2$ \;
	}
	\uElseIf{$e_1 < e_2$}{
		$\llbracket m_2 \rrbracket = \llbracket m_2 \rrbracket \otimes b^{e_2 - e_1}$ \;
		$e_n = e_1$ \;
	}
	\Else{
		$e_n = e_1$ \;
	}
	$\llbracket m_s \rrbracket = \llbracket m_1 \rrbracket \oplus \llbracket m_2 \rrbracket $ \;
	\Return \{$\llbracket m_s \rrbracket$, $e_n$\} \;
}
\caption{Paillier Floating Point Add}
\label{alg:PaillierFpAdd}
\end{algorithm}

A multiplication with a floating-point number that contains an encrypted mantissa ($\llbracket m_1 \rrbracket$, $e_1$) and a floating-point number with a plaintext mantissa ($m_2$, $e_2$) can be achieved by multiplying the mantissas with the multiplication operation defined for Paillier ($\llbracket m_n \rrbracket = \llbracket m_1 \rrbracket \otimes m_2 \;\;$\deleted{~ , \autoref{equ:paillierMultiplication}}) and a plaintext addition of the exponents ($e_n = e_1 + e_2 $).
This is also stated in line 10 and 11 of the \autoref{alg:PaillierFpMultiply}, which is sufficient for a correct result.
The lines from 2 to 9 contain a performance optimization, which prevents the intermediate result of $\llbracket m_e \rrbracket^{m_d}$, which is computed before $\mathrm{mod}\, N^2$ is applied in line 10, from being unnecessarily large\deleted{ (see \autoref{equ:paillierMultiplication})}.
This optimization is also used by the python library {\em python-paillier} \cite{web:PythonPaillierGit} in \texttt{paillier.py} 
and the java library {\em javallier} \cite{web:JavallierGit} in \texttt{PaillierContext.java}. 


\begin{algorithm}[tb]
\DontPrintSemicolon

\SetKwInOut{Param}{Parameters}
\SetKw{TypeInt}{integer}
\SetKwProg{Proc}{procedure}{}{}

\SetKwFunction{FnPfMultiply}{fpMultiply}

\BlankLine
\Param{Encrypted mantissa $\llbracket m_1 \rrbracket$, plaintext mantissa $m_2$ and the plaintext exponents ($e_1, e_2$) of the two floating point numbers that should be multiplied. \\ $N$ is the modulus of the used public key.}
\KwResult{Encrypt mantissa $\llbracket m_n \rrbracket$ and plaintext exponent $e_n$.}
\BlankLine

\Proc{\FnPfMultiply{$\llbracket m_1 \rrbracket$, $e_1$, $m_2$, $e_2$}}{
	$m_n = N - m_2$ \tcp*[h]{negative of $m_2$}\;

	\eIf(){$m_n \leq $ max. value that can be encrypted by current $N$} {
		$\llbracket m_e \rrbracket = \llbracket m_1 \rrbracket ^{-1} \mod N^2$
		\;
		$m_d = m_n$ \;
	} {
		$\llbracket m_e \rrbracket = \llbracket m_1 \rrbracket$ \;
		$m_d = m_2$ \;
	}

	$\llbracket m_n \rrbracket = \llbracket m_e \rrbracket \otimes m_d $ \;
	$e_n = e_1 + e_2 $ \;
	\Return \{$\llbracket m_n \rrbracket$, $e_n$\} \;
}
\caption{Paillier Floating Point Multiply}
\label{alg:PaillierFpMultiply}
\end{algorithm}

Signed numbers can be represented by using a two's complement representation for the mantissa $m$.
The exponent $e$ does not change.
If $v$ is a negative integer, the two's complement in the integer modulo $N$ can be calculated by: $m = v + N$.
In the encrypted domain, the additive inverse $-m$ of $m$ is defined by the multiplicitive inverse $\llbracket m \rrbracket ^{-1} = \llbracket i \rrbracket$ of $\llbracket m \rrbracket$ in the integers, modulo $N^2$
($\llbracket i \rrbracket$ is defined by: $\llbracket m \rrbracket \cdot \llbracket i \rrbracket = 1 \mod N^2$ and can be computed from $\llbracket m \rrbracket$ and $N^2$ by the {\em extended Euclidian algorithm} \cite{book:KnuthArtV2}).
This
complement representation for encrypted numbers can also be used for a subtraction of two encrypted numbers\deleted{, as already stated in \autoref{sec:PaillierCryptosystem}}.
Since, the first operand of a subtraction can be added to the additive inverse of the second operand ($\textrm{Dec}(\llbracket m_1 - m_2 \rrbracket) = \textrm{Dec}( \llbracket m_1 \rrbracket \times \llbracket m_2 \rrbracket^{-1} \: \mod \: N^2)$).

With the floating-point encoding explained in this section, it is possible to perform a trilinear interpolation of voxel values because the encrypted voxel values can be multiplied by the fractional distances between a sample position on a viewing ray and the actual voxel position.
Furthermore, divisions of an encrypted number $(\llbracket m \rrbracket, e)$ by a plaintext number $d$ can be approximated by a multiplication of the encrypted number $(\llbracket m \rrbracket, e)$ with the reciprocal $(\left\lfloor 1/d \cdot 10^\gamma \right\rceil , -\gamma)$ of $d$ ($\gamma$ defines the precision~ -~compare with \autoref{equ:fpDeviation}).

\section{Transfer Function}
\label{sec:transferFunction}
In this section, we discuss the challenges of building a transfer function approach that works for a probabilistic PHE scheme, and we show a novel and practical solution for a simplified transfer function.
It is not possible to use the transferred values for an alpha blending sample compositing because this would require a multiplication of two encrypted values, which is not possible with Paillier's cryptosystem.
However, the transfer function can be used to highlight specific density ranges at X-ray rendering, which helps an observer to distinguish between different objects inside a volume.

A transfer function for non-encrypted voxel values can be implemented as an array with the possible voxel values as indices and the assigned color as values of the array.
The evaluation of such a transfer function is as simple as reading the value from the array at the index, which is equal to the voxel value that should be mapped.
However, this cannot be efficiently implemented for encrypted data.
For non-encrypted voxel values, such a transfer function array will have a length that is equal to the amount of possible voxel values, which is only $2^8 = 256$ for $8$-bit voxels or $2^{10} = 1024$ for $10$-bit voxels.
An encrypted volume dataset will probably not contain two equal voxel values, because of the obfuscation during the encryption.
That means an encrypted dataset will probably have as many different voxel values as it has voxels.
Therefore,
an array as transfer function will not work because it would be at least as big as the volume itself.

Another approach for non-encrypted data is to store just some supporting points that contain the density and color.
The evaluation for this transfer function approach is achieved by interpolating the color between the value of the next lower and next greater supporting point.
To find the neighboring supporting points of the voxel value that should be transferred, comparison operators such as lower than ($<$) or greater than ($>$) are required.
However, comparison operators cannot exist for probabilistic PHE schemes like Paillier because that would break its security (see \autoref{sec:ComparisonOperators}).
Therefore, the question is how to implement a function $f:X \to Y$ that can map finite sets of numbers $X$ to another set of numbers $Y$ by just using the operations {\em add} ($\oplus$) and {\em multiply with constant} ($\otimes$).
The result of this function is again an encrypted number.
A promising approach that can achieve this was presented by Wamser et al. \cite{pdf:ObliviousLookupTables} in their work on \enquote{oblivious lookup-tables}.

\subsection{Oblivious Lookup Tables}
\label{sec:olut}
Let $X = \{x_1, x_2,...,x_n\}$ be an enumeration of values that should be mapped to $Y = \{y_1, y_2,...,y_n\}$ by the lookup function $f(x_i) = y_1$.
The idea is to create a vector $\vec{v_i}$ for every $x_i \in X$ with the same cardinality as $X$ ($|\vec{v_i}|=|X|$) and define the evaluation of a lookup by the dot product
$\vec{v_i} \cdot \vec{l} = y_i$.
The scalar value $y_i$ is the result of the lookup. For a transfer function, this would be the value of one color channel.
The vector $\vec{l}$ can be calculated form the linear
equation $V \cdot \vec{l} = \vec{y}$.
$V$ is a square matrix of full rank with $n = |X|$, that uses all vectors $\vec{v_i}$ as rows.
However, this linear equation needs to be solved only once.
Therefore, the client can calculate $\vec{l}$ upfront based on unencrypted numbers.
The
equation $V \cdot \vec{l} = \vec{y}$
has a unique solution, if all vectors  $\vec{v_i}$ are linearly independent.
Hence, the crucial part is to find an approach to extrapolate every vector $\vec{v_i}$ only from one single $x_i$ so that the $\vec{v_i}$ are linearly independent from each other.
Wamser et al. \cite{pdf:ObliviousLookupTables} suggest to use a Vandermonde-Matrix as $V$ (\autoref{equ:olutVandermonde}), because it fulfills these requirements.
\begin{align}
\begin{split}
	V = \begin{pmatrix}
	  1 & x_1^1 & x_1^2 & \cdots & x_1^{n-1} \\
	  1 & x_2^1 & x_2^2 & \cdots & x_2^{n-1} \\
	  \vdots  & \vdots & \vdots  & \ddots & \vdots  \\
	  1 & x_n^1 & x_n^2 & \cdots & x_n^{n-1}
	\end{pmatrix}
\end{split}
\label{equ:olutVandermonde}
\end{align}
From the creation rule of the Vandermonde-Matrix, it follows that a $\vec{v_i}$, which is equal to the $i$-th row of the matrix $V$, is defined as  $\vec{v_i} = (1, x_i^1, x_i^2, \cdots, x_i^{n-1})$.
The lookup function $f(x_i)$ can, therefore, be stated as:
\begin{align}
\begin{split}
	f(x_i) = (1, x_i^1, x_i^2, \cdots, x_i^{n-1}) \cdot \vec{l} = y_i
\end{split}
\label{equ:olutEvaluationVandermonde}
\end{align}
The dot product in \autoref{equ:olutEvaluationVandermonde}
can be calculated even if
$\vec{v_i} = (1, x_i^1, x_i^2, \cdots, x_i^{n-1})$
is encrypted because only the operations {\em add} ($\oplus$) and {\em multiply} ($\otimes$) that are defined for the Paillier HE are required for calculating a dot product.
However, it is not possible to calculate the vector $\vec{v_i}$ from an encrypted $\llbracket x_i \rrbracket$, because this would involve multiplications of two encrypted numbers, which is not possible with Paillier.
A theoretical solution for this could be to store the vector $\vec{v_i}$ instead of scalar $x_i$ as the value of a voxel.
For a volume dataset, where the voxel values have only a resolution of $8$ bits, this would lead to a vector length of $n = 2^8 = 256$.
Therefore, the required storage size for the volume will increase $256$ times.

A volume with $512 \times 512 \times 512$ voxels and a resolution of $8$ bits per voxel requires $512^3 \cdot 8~\mathrm{bits} / 8~\mathrm{bits} = 134,217,728~\mathrm{Bytes} = 128~\mathrm{MB}$.
The same volume encrypted by Paillier HE with a public key length that can be considered as secure ($2048$ bits) requires $512^3 \cdot 2 \cdot 2048 \mathrm{bits} / 8 \mathrm{bits}
= 64~\mathrm{GB}$.
If the scalar voxel values $x_i$ are replaced by the vectors $\vec{v_i}$ with a length of $256$, the volume will require $64~\mathrm{GB} * 256
= 16 \mathrm{TB}$.
While a volume dataset with $16$ Terabyte is probably better than a transfer function that is at least as big as the encrypted volume, the overhead in terms of storage and computation is still too big to be practical.
Therefore, we develop a simplified and novel transfer function approach with a considerably lower storage overhead, which we discuss in the next two sections.

\subsection{Density Range Emphasizing}
\begin{figure}[tb]
	\centering
	\subfigure[X-ray]{
		\includegraphics[width=0.478\columnwidth,trim=0 1mm 0 1mm,clip]{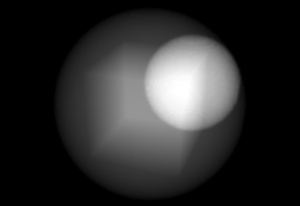}
	}
	\subfigure[emphasized density: 0.653]{
		\includegraphics[width=0.478\columnwidth,trim=0 1mm 0 1mm,clip]{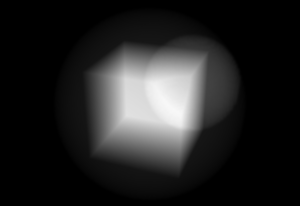}
	}
	\\
	\vspace*{-1mm}
	\subfigure[emphasized density: 0.331]{
		\includegraphics[width=0.478\columnwidth,trim=0 1mm 0 1mm,clip]{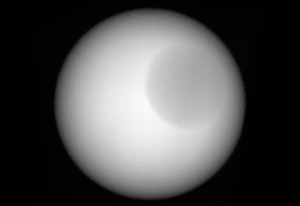}
	}
	\subfigure[emphasized density: 0.781]{
		\includegraphics[width=0.478\columnwidth,trim=0 1mm 0 1mm,clip]{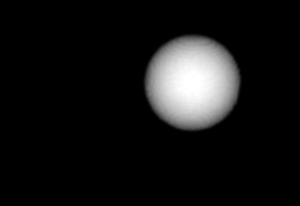}
	}
	\caption{First image shows an X-ray rendering result for comparison with the other three images that are created by our encrypted density emphasizing approach. The volume density values are encoded with 4-dimensional vectors.}
	\label{fig:densityEmphasising}
\end{figure}

Our simplified transfer function approach is based on the observation that it is possible to compute the dot product of a vector with encrypted values and a vector with plaintext values.
Furthermore, the dot product can be used to calculate an encrypted scalar value indicating  the similarity of an encrypted vector and a plaintext vector. This will work if both vectors have length $1$.
Therefore, our approach is to encode the density values of each voxel as a vector and encrypt each component of this vector by the Paillier encryption algorithm (Supplementary Material \autoref{alg:PaillierEncrypt}).
In order to highlight a user-defined density range, the density value at the center of this range needs to be encoded as a vector.
Note that this vector is not encrypted.
The encrypted volume rendering engine can now compute the dot product between this vector and the encrypted vector of a sample position.
Then the ray-casting algorithm needs to sum up the results of the dot products along a ray instead of the density values.
This approach allows a user to emphasize a selectable density range in the rendered image.
\autoref{fig:densityEmphasising} contains images that were created using this approach.
The top left subfigure shows a result of an X-ray rendering for comparison.
All other subfigures show results for different density ranges that are emphasized.
The density that is encoded as vector that was used for the dot-product calculation is specified in the caption of each sub figure.

\begin{figure}[tb]
	\centering 
	\subfigure{
		\includegraphics[width=0.478\columnwidth]{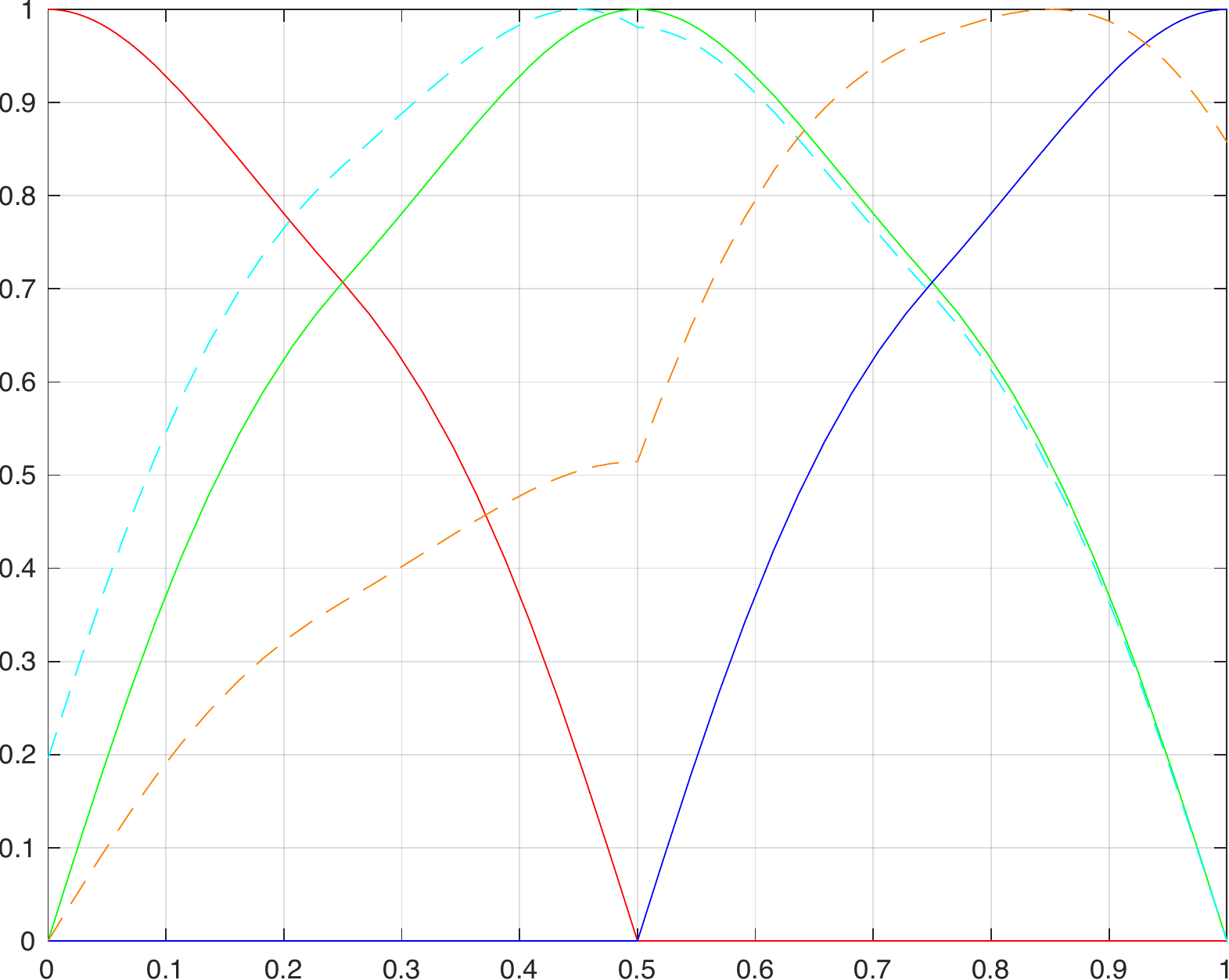}
	}
	\subfigure{
		\includegraphics[width=0.478\columnwidth]{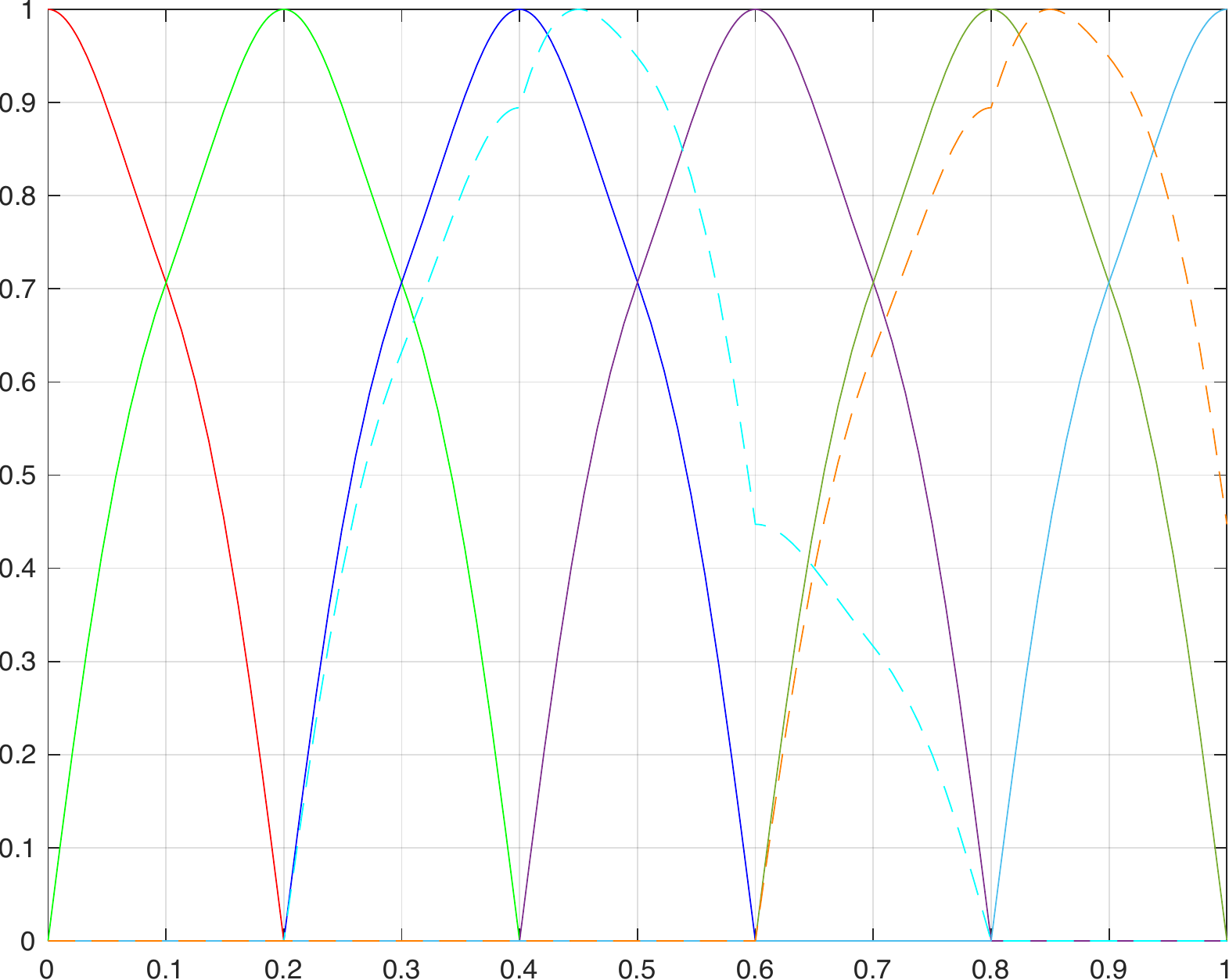}
	}
	\caption{
		Visualization of density encoded as 3 dimensional (left) and 6 dimensional (right) vectors.
		The scalar value (density) of the voxel is represented on the x-axes.
		The magnitude of each vector component at a specific density is represented by the curves.
		The first component is drawn in red, the second in green, red, purple, olive and light blue.
		The dashed curves shows the result of the dot product between the encoded voxel value and a {\em TF-Node} vector for a density of 0.45 in cyan and a density of 0.85 in orange.
	}
	\label{fig:densityEncodingPlots}
\end{figure}


\begin{algorithm}[tb]
\DontPrintSemicolon

\SetKwInOut{Param}{Parameters}
\SetKw{TypeInt}{integer}
\SetKwProg{Proc}{procedure}{}{}

\SetKwFunction{FnEncodeDensity}{encodeDensity}
\SetKwFunction{FnNormalize}{normalize}
\SetKwData{VarDensity}{density}
\SetKwData{VarDim}{dim}
\SetKwData{VarV}{v}

\BlankLine
\Param{The normalized \VarDensity that should be encoded as a vector with \VarDim dimensions.}
\KwResult{Vector \VarV}
\BlankLine

\Proc{\FnEncodeDensity{\VarDensity, \VarDim}}{
	initialize vector \VarV with length \VarDim and set all indices to $0$\;
	$s = \VarDensity \cdot 2 \cdot (\VarDim - 1)$\; 

	$f = (\lfloor s \rfloor + 1) / 2$ \; 
	$d = \lfloor f \rfloor$\; 

	$\VarV[d] = 1$ \;
	\uIf(){$d > 0$ {\bf and} $d = f $} {
		$\VarV[d-1] = 1 - (s - \lfloor s \rfloor)$ \;
	}
	\uElseIf{$d+1 < \VarDim$ {\bf and} $d < f $}{
		$\VarV[d+1] = s - \lfloor s \rfloor$ \;
	}

	\Return \FnNormalize(\VarV) \;
}
\caption{Encode Density}
\label{alg:encodeDensity}
\end{algorithm}

The density-to-vector encoding scheme we used is based on an HSV-to-RGB color conversion.
The exact encoding scheme is stated in \autoref{alg:encodeDensity}.
\autoref{fig:densityEncodingPlots} illustrates the magnitude of the vector components for all possible density values.
Furthermore, the response intensities for user-defined emphasizing densities at 0.45 and 0.85 are shown.
At the last line of \autoref{alg:encodeDensity}, the calculated vector is normalized.
This is important to make sure that the result of the dot product is always between $0$ and $1$ and to ensure that the highest possible dot product result ($1$) is at the user-defined emphasizing density.

There are other and possibly better density-to-vector encoding schemes.
However, the HSV-based encoding leads to results that feel natural, especially while smoothly increasing or decreasing the emphasizing density.
The encoding scheme should in any case be chosen in such a way that the curve created by the dot product is steep and narrow (see dashed lines in \autoref{fig:densityEncodingPlots}), so that the density selected by the user can be seen as clearly as possible in the resulting image.
The \autoref{alg:encodeDensity} takes not only the density that should be encoded as parameter, but also the count of dimensions of the returned vector.
Increasing the count of dimension not only makes the dot product response curve more steep
(See \autoref{fig:densityEncodingPlots} and compare the dashed lines in the left and right plot.),
but also increases the required storage size of the encoded and encrypted volume dataset.
Note that the count of dimensions must be the same during the encryption of the volume and for the encoding of the user-defined emphasizing density.
This also means that the amount of computations required for the volume rendering depends on the number of dimensions used for encoding the volume.

\subsection{Simplified Transfer Function}
\begin{figure}[tb]
	\centering
	\subfigure[blue at 0.279, red at 0.797]{
		\includegraphics[width=0.478\columnwidth,trim=0 1mm 0 1mm,clip]{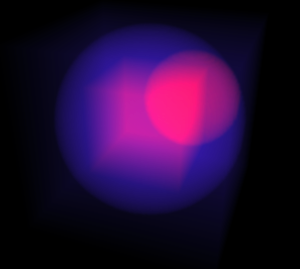}
		\label{fig:generic4dimColored_xray}
	}
	\subfigure[blue at 0.000, red at 1.000]{
		\includegraphics[width=0.478\columnwidth,trim=0 1mm 0 1mm,clip]{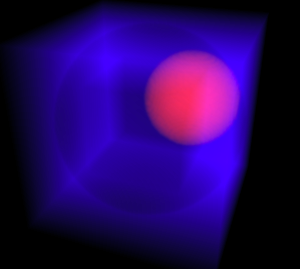}
	}
	\\
	\vspace*{-1mm}
	\subfigure[green at 0.076, blue at 0.651, red at 1.000]{
		\includegraphics[width=0.478\columnwidth,trim=0 1mm 0 1mm,clip]{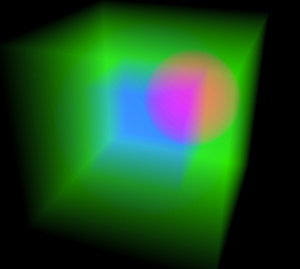}
	}
	\subfigure[blue at 0.000, yellow at 0.293, green at 0.664, purple at 1.000]{
		\includegraphics[width=0.478\columnwidth,trim=0 1mm 0 1mm,clip]{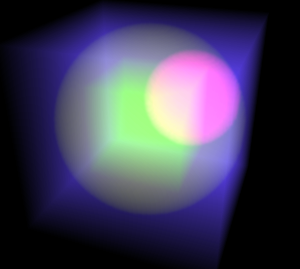}
	}
	\caption{Images are created by our simplified transfer function approach. The volume data voxel values are encoded by four-dimensional vectors. The subfigures shows results of different transfer functions applied to the same encrypted dataset.}
	\label{fig:generic4dimColored}
\end{figure}

It is possible to add RGB colors to the rendered images based on the density range emphasizing described in the last section.
This is useful because RGB colors allow a user to emphasize different densities in the same image while keeping the densities distinguishable (see \autoref{fig:generic4dimColored}).
Since the dot product between an encoded and encrypted voxel value and a user-defined encoded density is an encrypted scalar value, a multiplication with another plaintext number is possible.
For our simplified transfer function approach, the dot product result needs to be multiplied with a user-defined RGB color vector.
As the dot product expresses the similarity between the voxel value and the user-defined density, the intensity of the resulting RGB color will be high if the densities are similar, and low otherwise.
Since the RGB color vector is not encrypted,
the multiplication between the encrypted dot product result and the RGB color vector can be archived by three separate homomorphic multiplications ($\otimes$) of one encrypted and one plaintext number\deleted{ (see \autoref{equ:paillierMultiplication})}.
The result of such a multiplication is an encrypted RGB color.
This calculation can be performed not only for one density-RGB-color-pair, but also for multiple such pairs.
For a better understanding, we will call such a pair consisting of a density and an RGB color a {\em transfer function node (TF-Node)}.

\autoref{equ:encVoxel2Rgb} shows the transformation for one encoded and encrypted voxel value $\llbracket \vec{v} \rrbracket$ to an encrypted RGB color $\llbracket \vec{c_v} \rrbracket$.
The symbol $\bigoplus$ is used instead of $\sum$, because the sum of encrypted vectors needs to be calculated.
The variable $n$ denotes the count of user defined {\em TF-Nodes}.
The vectors $\vec{d_i}$ and $\vec{c_i}$ are the encoded density and RGB color of the {\em TF-Node} with index $i$.
The symbol $\odot$ is used as operator for a dot product between one encrypted vector and one plaintext vector.

\begin{align}
\begin{split}
	\llbracket \vec{c_v} \rrbracket = \bigoplus_{i=0}^{n}\left( \llbracket \vec{v} \rrbracket \odot \vec{d_i} \right) \otimes \vec{c_i}
\end{split}
\label{equ:encVoxel2Rgb}
\end{align}
To obtain the final encrypted RGB color of a pixel, the sum of all encrypted RGB sample values $\llbracket \vec{c_v} \rrbracket$ along a viewing ray needs to be calculated.
The total RGB vector needs to be divided by the sample count as usual for averaging and, furthermore,
by the count of {\em TF-Nodes}.
This can be achieved by dividing each component of the total RGB vector by the product of the sample count and the count of {\em TF-Nodes}.
The method to approximate a division of an encrypted number is stated in \autoref{equ:fpDeviation}.
After calculating this for every image pixel, the entire encrypted image is sent to the client.
A client that knows the right secure key can now decrypt each RGB component of each pixel and display the colored image.
Example images rendered with this approach are shown in \autoref{fig:teaser} and \autoref{fig:generic4dimColored}.

\section{Results}
\label{sec:Results}
All performance tests are executed on a Mac Book Pro (15-inch, 2016) with an 2.9 GHz Intel Core i7.
All algorithms are implemented in Java \added{and are only single-threaded}.
The purpose of the implementation is to prove the concept and, in its current form, is not performance-optimized.
All runtimes shown in \autoref{table:performanceXray} and \autoref{table:performanceColor} are measured with volume size of $100 \times 100 \times 100$ voxels.
The rendered image always has a size of $150 \times 150$ pixels.

\autoref{table:performanceXray} shows the runtime performance required for encrypting a volume with scalar voxel values, X-ray rendering, and image decryption with different public key modulus lengths. The table is divided into four groups of rows.
The first two groups show the required time for rendering with nearest-neighbor sampling.
Group three and four show the resulting performance for trilinear interpolation.
The numbers in group one and three of the \autoref{table:performanceXray} are measured without obfuscation during the encryption; therefore, the encrypted volume is not secure.
While this type of \enquote{encryption} does not have any practical relevance, it is interesting to compare these runtime numbers with those in the group two and four, which are measured from a secure encryption with obfuscation.
It can be seen that the obfuscation takes a significant amount of time.
Therefore, the random number generation ($r$) that is required for the obfuscation and the calculation of $r^N$ (see Supplementary Material \autoref{alg:PaillierEncrypt}) has a substantial impact on the time required for encrypting the volume dataset.
We use the \texttt{java.security.SecureRandom} class from the java standard runtime framework as random number generator for the obfuscation.

\newcommand{\STAB}[1]{\begin{tabular}{@{}c@{}}#1\end{tabular}}

\bgroup
\def\arraystretch{1.5}%
\begin{table}[tb]
\caption{X-ray: Required time (in seconds) for encryption, rendering and decryption with different modulus lengths.}
\label{table:performanceXray}
\resizebox{\columnwidth}{!}{%
	\begin{tabular}{l|l|l||r|rrrrrr}
		 \multicolumn{1}{c}{} & \multicolumn{1}{c}{} & & \multicolumn{1}{c|}{\textbf{plain}}	& \multicolumn{1}{c}{\textbf{64bit}} & \multicolumn{1}{c}{\textbf{128bit}} & \multicolumn{1}{c}{\textbf{256bit}} & \multicolumn{1}{c}{\textbf{512bit}} & \multicolumn{1}{c}{\textbf{1024bit}} & \multicolumn{1}{c}{\textbf{2048bit}} \\
		\hhline{===||=|======}
	%
	\multirow{6}{*}{\STAB{\rotatebox[origin=c]{90}{\shortstack{nearest \\ neighbor}}}}
	& \multirow{3}{*}{\STAB{\rotatebox[origin=c]{90}{\shortstack{without \\ obfuscation}}}}
	& \textbf{encrypt}	&					& 0.46		& 0.48		& 0.54 		& 0.56 		& 0.62		& 0.57 \\
	& & \textbf{render}		& 0.03		& 0.54 		& 0.66		& 0.89		& 1.61		& 3.49		& 9.49		\\
	& & \textbf{decrypt}	&					& 0.17		& 0.25		& 0.63		& 2.56		& 14.92	& 99.56	\\
	\cline{2-10}
	%
	& \multirow{3}{*}{\STAB{\rotatebox[origin=c]{90}{\shortstack{with \\ obfuscation}}}}
	& \textbf{encrypt}	&					& 5.72		& 15.36		& 59.54 		& 327.24 		& 2256.01		& 16880.00 \\
	& & \textbf{render}		& 0.03		& 1.10 		& 1.77		& 4.18		& 11.61		& 37.10		& 94.30		\\
	& & \textbf{decrypt}	&					& 0.21		& 0.43		& 1.36		& 4.94		& 27.24	& 185.94	\\
	\cline{1-10}
	%
	\multirow{6}{*}{\STAB{\rotatebox[origin=c]{90}{\shortstack{trilinear \\ interpolation}}}}
	& \multirow{3}{*}{\STAB{\rotatebox[origin=c]{90}{\shortstack{without \\ obfuscation}}}}
	& \textbf{encrypt}	& 				& 0.46		& 0.47		& 0.52		& 0.57		& 0.54		& 	0.65	\\
	& & \textbf{render}   & 0.08		& 10.59		& 13.55		& 21.38		& 47.25		& 146.07		& 	487.58 \\
	& & \textbf{decrypt}	&     		& 0.14		& 0.26		& 0.64			& 2.59			& 14.65		&		100.72	\\
	\cline{2-10}
	%
	& \multirow{3}{*}{\STAB{\rotatebox[origin=c]{90}{\shortstack{with \\ obfuscation}}}}
	& \textbf{encrypt}	& 				& 5.82		& 14.67		& 59.93		& 330.56		& 2226.01		& 	16512.47	\\
	& & \textbf{render}   & 0.08		& 16.67		& 23.86		& 47.07		& 121.05		& 385.71		& 	1182.48	\\
	& & \textbf{decrypt}	&     		& 0.20		& 0.41		& 1.20			& 4.89			& 26.86		&		186.38	\\
	\cline{1-10}
	\end{tabular}
}
\end{table}
\egroup

\bgroup
\def\arraystretch{1.5}%
\begin{table}[tb]
\caption{
	Required storage size for an encrypted volume with $100 \times 100 \times 100$ voxels and different modulus lengths.
}
\label{table:volumeStorageSize}
\resizebox{\columnwidth}{!}{%
	\begin{tabular}{c||r|rrrrrr}
		& \multicolumn{1}{c|}{\textbf{plain (8bit)}}	& \multicolumn{1}{c}{\textbf{64bit}} & \multicolumn{1}{c}{\textbf{128bit}} & \multicolumn{1}{c}{\textbf{256bit}} & \multicolumn{1}{c}{\textbf{512bit}} & \multicolumn{1}{c}{\textbf{1024bit}} & \multicolumn{1}{c}{\textbf{2048bit}} \\
		\hhline{=||=|======}
		scalar &
		1 MB &
		16 MB &
		32 MB &
		64 MB &
		128 MB &
		256 MB &
		512 MB  \\
		\cline{1-8}
		2 dim &
		2 MB &
		32 MB &
		64 MB &
		128 MB &
		256 MB &
		512 MB &
		1024 MB  \\
		\cline{1-8}
		3 dim &
		3 MB &
		48 MB &
		96 MB &
		192 MB &
		384 MB &
		768 MB &
		1536 MB  \\
		\cline{1-8}
		4 dim &
		4 MB &
		64 MB &
		128 MB &
		256 MB &
		512 MB &
		1024 MB &
		2048 MB  \\
		\cline{1-8}
	\end{tabular}
}
\end{table}
\egroup

\autoref{table:volumeStorageSize} shows the required memory size for this volume with a single scalar value per voxel and also for encodings in multiple dimensions at different modulus lengths.
\autoref{table:performanceColor} shows the runtime required for encrypting a volume with different voxel encodings (two, three, and four dimensions), rendering with our simplified transfer function approach at different counts of {\em TF-Nodes} (one, two, ... colors) and image decryption. The resulting performance for all these operations is provided for different public key modulus lengths.

\bgroup
\def\arraystretch{1.5}%
\begin{table}[tb]
\caption{Simplified transfer function: required time (in seconds) for encryption, rendering and decryption with different modulus lengths.}
\label{table:performanceColor}
\resizebox{\columnwidth}{!}{%
	\begin{tabular}{l|l|l||r|rrrrrr}
		 \multicolumn{1}{c}{} & \multicolumn{1}{c}{} & & \multicolumn{1}{c|}{\textbf{plain}} & \multicolumn{1}{c}{\textbf{128bit}} & \multicolumn{1}{c}{\textbf{256bit}} & \multicolumn{1}{c}{\textbf{512bit}} & \multicolumn{1}{c}{\textbf{1024bit}} & \multicolumn{1}{c}{\textbf{2048bit}} \\
		\hhline{===||=|======}
	\multirow{5}{*}{\STAB{\rotatebox[origin=c]{90}{\shortstack{2 dimensional \\ encoding}}}}
	& & \textbf{encrypt}	&					& 29.87		& 115.03 	& 622.77 		& 4405.69		& 31796.00 \\
	\cline{2-10}
	& \multirow{2}{*}{\STAB{\rotatebox[origin=c]{90}{\shortstack{1 {\small color}}}}}
	& \textbf{render}			& 0.05		& 21.85		& 48.59		& 142.36		& 544.35		& 1999.51		\\
	& & \textbf{decrypt}	&					& 0.80		& 1.37		& 4.95			& 27.7			& 180.40	\\
	\cline{2-10}
	& \multirow{2}{*}{\STAB{\rotatebox[origin=c]{90}{\shortstack{2 {\small colors}}}}}
	& \textbf{render}			& 0.06		& 23.38		& 58.20		& 131.10	& 659.48			& 2716.51		\\
	& & \textbf{decrypt}	&					& 0.33		& 1.18		& 3.23		& 27.09				& 187.51	\\
	\hhline{=|=|=||=|======}
	\multirow{7}{*}{\STAB{\rotatebox[origin=c]{90}{\shortstack{3 dimensional \\ encoding}}}}
	& & \textbf{encrypt}	& 				& 40.30		& 162.22	& 897.28		& 6271.48		& 47030.98	\\
	\cline{2-10}
	& \multirow{2}{*}{\STAB{\rotatebox[origin=c]{90}{\shortstack{1 {\small color}}}}}
	& \textbf{render}   	& 0.05		& 19.59		& 40.21		& 143.69		& 560.98		& 2299.65	\\
	& & \textbf{decrypt}	&     		& 0.53		& 0.83		& 4.47			& 25.98			&	127.84	\\
	\cline{2-10}
	& \multirow{2}{*}{\STAB{\rotatebox[origin=c]{90}{\shortstack{2 {\small colors}}}}}
	& \textbf{render}   	& 0.06		& 25.87		& 56.24		& 154.71		& 770.64		& 2850.69	\\
	& & \textbf{decrypt}	&     		& 0.43		& 0.80		& 3.18			& 27.67			&	120.99	\\
	\cline{2-10}
	& \multirow{2}{*}{\STAB{\rotatebox[origin=c]{90}{\shortstack{3 {\small colors}}}}}
	& \textbf{render}   	& 0.06		& 41.67		& 89.73		& 257.51		& 919.41		& 3685.33	\\
	& & \textbf{decrypt}	&     		& 0.59		& 1.21		& 4.83			& 25.46			&	181.86	\\
	\hhline{=|=|=||=|======}
	\multirow{9}{*}{\STAB{\rotatebox[origin=c]{90}{\shortstack{4 dimensional \\ encoding}}}}
	& & \textbf{encrypt}	& 				& 54.97		& 213.21	& 1199.68		& 8512.23		& 62959.05	\\
	\cline{2-10}
	& \multirow{2}{*}{\STAB{\rotatebox[origin=c]{90}{\shortstack{1 {\small color}}}}}
	& \textbf{render}   	& 0.06		& 19.46		& 42.93		& 98.08			& 429.30		& 2292.48	\\
	& & \textbf{decrypt}	&     		& 0.28		& 0.77		& 1.54			& 8.50			&	60.03	\\
	\cline{2-10}
	& \multirow{2}{*}{\STAB{\rotatebox[origin=c]{90}{\shortstack{2 {\small colors}}}}}
	& \textbf{render}   	& 0.07		& 35.41		& 61.60		& 179.45		& 729.11		& 3234.38	\\
	& & \textbf{decrypt}	&     		& 0.46		& 0.76		& 3.10			& 18.36			&	121.76	\\
	\cline{2-10}
	& \multirow{2}{*}{\STAB{\rotatebox[origin=c]{90}{\shortstack{3 {\small colors}}}}}
	& \textbf{render}   	& 0.07		& 46.41		& 93.63		& 268.59		& 1045.56		& 4251.69	\\
	& & \textbf{decrypt}	&     		& 0.45		& 1.11		& 4.52			& 26.15			&	190.63-	\\
	\cline{2-10}
	& \multirow{2}{*}{\STAB{\rotatebox[origin=c]{90}{\shortstack{4 {\small colors}}}}}
	& \textbf{render}   	& 0.08		& 63.57		& 123.32	& 343.20		& 1320.86		& 5217.38	\\
	& & \textbf{decrypt}	&     		& 0.67		& 1.51		& 4.68			& 26.82			&	190.49	\\
	\hhline{==========}
	\end{tabular}
}
\end{table}
\egroup

\added{
	The rendering results of \autoref{fig:teaser} show what can be done with our simplified transfer function.
	The right image demonstrates the  utilization in nuclear medicine.
	During the diagnosis, these datasets are usually investigated either by showing single slices or by X-ray renderings, where the depth cues are provided through rotating the dataset around an axis. This  is possible with our homomorphic-encrypted volume rendering with the added  privacy, which is useful for diagnosing from such a highly sensitive type of modality and associated pathologies.
}

\section{Discussion}
\added{
	First, we discuss possible performance improvements of our prototype and how the approach could scale to interactive frame rates for larger real-world datasets.
	Later, starting with general noteworthy considerations, we discuss security-related aspects of our volume rendering approach. Finally, we follow with an explanation for the invisibility of comparisons.
	In \autoref{sec:FPNumbersSecurity}, we will show that the used floating-point encoding with an encrypted mantissa and a plaintext exponent does not weaken the privacy of the encrypted volume data.
}

\subsection{\added{Performance}}%
\label{sec:PerformanceImprovements}%
Our prototype is implemented as a single-threaded application; however, a major strength of our approach is that it is highly parallelizable and should scale linearly with the processing power.
There are obvious opportunities to improve the performance to a multi-threaded implementation, and multiple memory allocations (\texttt{new} statements) during the rendering could be avoided.
During the encryption, every voxel can be processed independently.
Therefore, it should be relatively easy to use as many processing units (e.g., CPU cores or shader hardware on GPU) as voxels in the volume for the encryption.
In the rendering and decryption stage, every pixel of the image can be processed independently.
Therefore, the number of processing units that can be used efficiently in parallel is equal to the number of pixels in the final image.
Furthermore, a better storage order of voxel values, such as Morton order \cite{tech:mortonOder} (recursive Z curve) extended to three dimensions, could lead to a better cache usage, which will further improve the performance.
The implementation used for all shown results is based only on a naive three-dimensional \texttt{BigInteger} array as volume storage.

\added{
If we consider a real-world dataset with a resolution of $512 \times 512 \times 512$ voxels encrypted with a perfectly secure 2048bit long key for the purpose of X-ray rendering
with a single value per voxel, the encrypted dataset will have a size of 64GB ($=(512^3 \cdot 2048 \cdot 2) / (8 \cdot 1024^3)$).
While this is a considerable data amplification compared to the 16bit plaintext representation of the dataset with $256$MB ($=(512^3 \cdot 16) / (8 \cdot 1024^2)$), it will nevertheless perfectly fit in the video-memory of two NVIDIA Quadro RTX 8000 that have $48$~GB of 
memory each.
An encrypted volume with a four-dimensional encoding for our simplified transfer function approach will be four times bigger and will, therefore, have a size of $256$GB.
Consequently, at least six GPUs with $48$GB memory each will be required.
While six GPUs in one server is absolutely possible, our privacy-preserving volume rendering approach should scale much further.
It should be possible to use our proposed encrypted voxel compositing scheme as mapper for the MapReduce implementation proposed by Stuart et al. \cite{inProc:GPUMapReduce}, which can make use of a GPU-accelerated distributed memory system for volume rendering.
}

\subsection{\added{Security Considerations}}
The data privacy of our approach depends entirely on the security of Paillier's cryptosystem.
Our approach does not store any voxel value or any information that is computed from a voxel value without an encryption by Paillier's cryptosystem.
The Paillier cryptosystem is semantically secure against chosen-plaintext attacks (IND-CPA) \cite{inbook:homoEncApps:2014:paillier}.
Therefore, we conclude that the data that our approach provides to the storage and rendering server are protected in a semantically secure way.
The computational complexity required for breaking a secure key of Paillier's cryptosystem depends on the length of the modulus $N$.
The larger the modulus $N$ is, the harder it is to be factorized, which would be required for data decryption.
For the required length of the modulus, the same conditions as for the RSA cryptosystem \cite{article:rsa} should hold.
From 2018 until 2022, a modulus $N$ with a length of at least 2048 bits is considered to be secure \cite{web:cryptographicKeyLength, NIST:800_56B_Rev.2}.

\subsection{Encrypted Comparison Operators}
\label{sec:ComparisonOperators}
It is not possible to compare encrypted numbers with each other.
During the encryption of a number, the obfuscation is performed,
\deleted{(see \autoref{sec:PaillierCryptosystem})}
which randomly distributes the encrypted values between $0$ an $N^2-1$.
Therefore, the order of the encrypted values $\llbracket M \rrbracket$ has nothing to do with the order of the underlying numbers $M$ that were encrypted.
Consequently, operators such as lower than ($<$) or greater than ($>$) cannot provide a result that is meaningful for the numbers $M$, if they are applied to encrypted values $\llbracket M \rrbracket$.

We can also argue that comparison operators cannot exist if the Paillier cryptosystem is secure, since the existence of a comparison operator would break the security of the cryptosystem.
Consider a less-than comparison for example: if such a comparator could be implemented, every value could be decrypted within $\mathrm{log_2}(N)$ comparisons by a binary search.
For a modulus $N$ with a length of $2048$ bit, an attacker would need to encrypt and then compare only  $\mathrm{log_2}(2^{2048}) = 2048$ numbers with the encrypted value $\llbracket m \rrbracket$ in order to find the decrypted number $m$.
This would effectively break the security of the encryption scheme.

\subsection{Plaintext Exponent Does Not Leak Private Data}
\label{sec:FPNumbersSecurity}
At first glance, it may look like the floating-point representation (encrypted mantissa, plaintext exponent) we used will allow an attacker to obtain more important information than within an encoding where all number components are encrypted.
However, if it is implemented correctly, an attacker cannot take any advantage from this number representation.
\added{
	First, we will discuss this for the data in the server memory and, in the last paragraph, we will show how the exponent can be protected during the data transfer from the server to the client.
}

For the following, we will suppose a secure system with an at least 2048-bit long modulus $N$ and, therefore, a mantissa $\llbracket m \rrbracket$ with at least $600$ decimal digits usable in the plaintext domain.
Voxel values that are stored as 10 bit values are probably precise enough for most volume-rendering use cases.
To store numbers between $0$ and $2^{10} = 1024$, the exponent $e$ is not required at all, because the voxel information can be stored only in the mantissa $m$.
Therefore, the exponent $e$ can be $1$ for all voxels.
This means that the exponent does not even have to be transferred to the server, because the server can implicitly assume that the exponents of all numbers is $1$.
An addition of any of these numbers that have an exponent of $1$ does not change the exponent, because for an addition, the exponent needs to be taken into account only if the summands have different exponents (see \autoref{alg:PaillierFpAdd}).
Therefore, only a multiplication (e.g., an interpolation between voxel values) can change the exponent to anything other than $1$.
However, the Paillier cryptosystem only supports the multiplication of an encrypted number with an unencrypted number.
Consequently, the number $d$ that changes an exponent has to be unencrypted.
Furthermore, this number $d$ can only depend on unencrypted data, because Paillier does not support comparison operators (see \autoref{sec:ComparisonOperators}), which are required for flow control statements like \texttt{if} or \texttt{for-loops}, and arithmetic operations with an encrypted number will result in useless random noise, except those add ($\oplus$) and multiply ($\otimes$) that are defined for the Paillier cryptosystem%
.
Therefore, the number $d$ can only be the result of some computation with other unencrypted variables.
This implies that $d$ does not need to be encrypted, because everyone can calculate $d$ itself.
In other words, if the variable $d$ can be computed from some variables that need to be considered as publicly available, because they are unencrypted, it is pointless to encrypt $d$.
If $d$, which is unencrypted and can only depend on unencrypted data, influences an exponent, the exponent exposes only the information that is already publicly available.

The important observation here is that an unencrypted value (e.g., an exponent) can influence an encrypted value (e.g., a mantissa), but an encrypted value (e.g., a mantissa) cannot influence an unencrypted value (e.g., an exponent).
This means that no information that is only available as encrypted data can ever be exposed in unencrypted values like the exponent.

In our rendering system, a number $d$ that changes an exponent can either be the result of a computation with a constant or with an unencrypted number that is provided in unencrypted form to the rendering system, such as the camera properties (position of eye point, opening angle, view direction...).
Therefore, an attacker could possibly learn the constants used in our program code and data, such as the camera properties that are provided in the unencrypted form, from the exponents of the rendering result (the image).
However, we want to develop an approach that is open and semantically secure
\added{by design and not {\em secure through obscurity}}
(compare: \cite{article:hoepman2008securityThrough, NIST:800_123, web:BruceSchneierTheInsecurity2, web:ChadPerrinSecurity101}).
Therefore, we have to treat the source code of the application as publicly available, which means that a constant cannot be considered to be private.
Furthermore, for our approach, the camera properties need to be provided in an unencrypted form to the rendering system.
Therefore, we cannot consider it as private anyway.

It should be noted that the camera properties could possibly provide interesting information to an attacker, because it could be possible to learn something about the volume data by tracking the camera properties over time.
For instance, if a user rotates the camera around a specific region for a considerable amount of time, an attacker could guess that the region contains some interesting data.
During the transfer of the camera properties from the client to the server over the network, the camera properties could be secured by using an encrypted tunnel, such as {\em IPsec} \cite{tech:RFC4301.IPsec} or {\em TLS} \cite{tech:RFC8446.TLS}.
However, our basic assumption is that we cannot trust the server that hosts our rendering program.
This means that an attacker has access to the entire memory of the server and, therefore, can read the camera properties directly from the memory of the server, regardless of the used network transfer method.
While the unencrypted camera properties could
\replaced{indirectly expose some information}{be a security problem}, we will not discuss this further because it is beyond the scope of this work.

Based on the arguments stated in this section, we can conclude that using plaintext exponents for the rendering process on an untrusted computer system does not provide more information to a third party than using encryption for all components of a floating-point number.

The only remaining part that needs to be considered is the transfer of the final image from the server back to the client across a network.
Operations like trilinear interpolation will change the exponents during the rendering.
Therefore, the final image will contain floating-point numbers with exponents unequal to $1$ and, because the interpolation weights that change the exponents depend on the camera properties, the exponents of the final image will provide some information about the camera properties.
The privacy of the information that is stored in the exponents is only important if it can be assumed that the server is trustworthy, which contradicts the basic assumption of this work.
Therefore, this is somewhat beyond the scope of this work, but we nonetheless discuss it here for the sake of completeness.
In order to encrypt as much information as possible during the image transfer from the server to the client, ideally all information should be stored in the encrypted mantissa.
While it is not possible to divide an encrypted number, it is possible to multiply an encrypted number.
Furthermore, the encrypted mantissa can store numbers in the range from $0$ to $2^{2047}$.
Therefore, it is possible to bring all exponents to the value of the smallest exponent of any pixel of the final image.
This can be achieved by the calculation shown in \autoref{equ:decreaseExponentTo}.
For the new exponent $e_n$, the value of the smallest exponent of any pixel must be used.
If this exponent-decrease operation is applied to all image values on the server before transferring the image to the client, the exponent should not contain any important information during the transfer, because all exponents then contain the same value.
However, if there is concern that even this might contain something useful, it is possible to encrypt this exponent with the public key because the client that has the secure key can decrypt it anyway.
Since it is the same value for every number that is sent back to the client, this exponent needs to be sent and decrypted only once.

\section{Conclusions}
While the expressiveness of our renderings is far from what is possible with state-of-the-art algorithms for non-encrypted data, we have presented a highly parallelizable direct volume rendering approach that allows not only the outsourcing of the storage of the volume data, but also the outsourcing of the whole rendering pipeline, without compromising the privacy of the data.
The approach we propose does not leak any voxel values or any information computed from a voxel value after the volume encryption.
Since we encrypt every single bit of voxel data with Paillier's cryptosystem, which is provably semantically secure (see: \cite{pdf:Paillier, inbook:homoEncApps:2014:paillier}), it is rather obvious that with our approach, the confidentiality of the volume data (densities, shapes, structures,..) and the colors of the rendered image only depends on the privacy of the secure key.
If we trust all devices that have seen the volume data before encryption (e.g.,: MRI-/CT-scanner, the computer that performs the encryption) to safely delete the data after encryption, only the owner of the secure key is able to obtain any useful information of the encrypted volume or rendered images.
This is a significant advantage compared to all previous works to date.

This security naturally comes with associated costs. The storage overhead costs for computation are between four and five orders of magnitude compared to plaintext data.
Compared to our prototype, an optimized implementation of our approach can reduce the computational complexity by an order of magnitude.

While we hope that further improvements of our approach would lead to rendering results with better expressiveness, it will be a non-trivial task because the security aspect needs to be considered for even the slightest change.
Many of the ideas we considered in the algorithmic design eventually led to a leak of sensitive information, which is, in our opinion, intolerable, no matter how small it may be.
\added{Future work definitively needs to improve the rendering performance. We see that the performance can be tremendously accelerated, as ray-casting is an \emph{embarrassingly parallel} workload.
For practical utilization of our privacy-preserving volume rendering, an efficient GPU-based implementation would be necessary.
	A single server full with GPUs should be able to provide five orders of magnitude more computational power than a single CPU core can.
	Based on the measured performance with a non-optimized single threaded implementation, such a server could be able to achieve interactive frame rates for datasets that are small enough to fit into the memory of the graphics cards.
Therefore, we see, as a next step, to port the rendering onto GPUs, where the necessary technological piece will be to design efficient big-integer arithmetic. Another possible improvement within the scope of Paillier HE will be the visual quality of compositing. This can be done with gradient-magnitude opacity modulation, where the gradient magnitude will be pre-calculated and encrypted along with the data values. Such representation can already lead to substantial visual quality improvement, although it will still not reach the outcome of compositing using Porter/Duffs's over operator~\cite{Porter1984}. For the Paillier HE scheme, we do not see a way to implement the over operator compositing, as it requires a multiplication of encrypted numbers. To support alpha blending, new research should be oriented on investigating other homomorphic encryption schemes or a combination of those that, unlike Paillier, would support desired secure alpha blending functionality.}

\acknowledgments{
The authors wish to thank Michal Hojsík for his fruitful discussions on cryptography.
The authors would like to thank Michael Cusack from Publication Services at KAUST for proofreading.
The research was supported by King Abdullah University of Science and Technology (KAUST) under award number BAS/1/1680-01-01.}
\bibliographystyle{abbrv-doi-hyperref}

\bibliography{template}

\begin{thebibliography}{10}

\bibitem{article:AbbasHeSurvey2018}
\href{https://doi.org/10.1145/3214303}{A.~Acar, H.~Aksu, A.~S. Uluagac, and
  M.~Conti}.
\newblock \href{https://doi.org/10.1145/3214303}{A survey on homomorphic
  encryption schemes: Theory and implementation}.
\newblock \href{https://doi.org/10.1145/3214303}{{\em ACM Computing Surveys}},
  \href{https://doi.org/10.1145/3214303}{51(4):79:2--79:35},
  \href{https://doi.org/10.1145/3214303}{2018}.
  \href{https://doi.org/10.1145/3214303}
{doi: {{%
10\hspace{.1pt}\discretionary{.}{%
}{.}\hspace{.4pt}1145\discretionary{/}{%
}{/}3214303}}}


\bibitem{NIST:800_56B_Rev.2}
\href{https://doi.org/10.6028/NIST.SP.800-56Br2}{E.~B. Barker, L.~Chen, A.~L.
  Roginsky, and R.~D.~S. Simon}.
\newblock \href{https://doi.org/10.6028/NIST.SP.800-56Br2}{Recommendation for
  pair-wise key establishment using integer factorization cryptography}.
\newblock \href{https://doi.org/10.6028/NIST.SP.800-56Br2}{Technical Report
  NIST Special Publication 800-56B Rev. 2},
  \href{https://doi.org/10.6028/NIST.SP.800-56Br2}{U.S. Department of Commerce,
  National Institute of Standards and Technology, Gaithersburg, MD},
  \href{https://doi.org/10.6028/NIST.SP.800-56Br2}{2019}.
  \href{https://doi.org/10.6028/NIST.SP.800-56Br2}
{doi: {{%
10\hspace{.1pt}\discretionary{.}{%
}{.}\hspace{.4pt}6028\discretionary{/}{%
}{/}NIST\hspace{.1pt}\discretionary{.}{%
}{.}\hspace{.4pt}SP\hspace{.1pt}\discretionary{.}{%
}{.}\hspace{.4pt}800\discretionary{%
}{-}{-}56Br2}}}


\bibitem{ObfuscatedVolRend}
\href{https://doi.org/10.1007/s00371-015-1143-6}{J.-K. Chou and C.-K. Yang}.
\newblock \href{https://doi.org/10.1007/s00371-015-1143-6}{Obfuscated volume
  rendering}.
\newblock \href{https://doi.org/10.1007/s00371-015-1143-6}{{\em The Visual
  Computer}},
  \href{https://doi.org/10.1007/s00371-015-1143-6}{32(12):1593--1604},
  \href{https://doi.org/10.1007/s00371-015-1143-6}{2016}.
  \href{https://doi.org/10.1007/s00371-015-1143-6}
{doi: {{%
10\hspace{.1pt}\discretionary{.}{%
}{.}\hspace{.4pt}1007\discretionary{/}{%
}{/}s00371\discretionary{%
}{-}{-}015\discretionary{%
}{-}{-}1143\discretionary{%
}{-}{-}6}}}


\bibitem{article:datasetTCIA}
\href{https://doi.org/10.1007/s10278-013-9622-7}{K.~Clark, B.~Vendt, K.~Smith,
  J.~Freymann, J.~Kirby, P.~Koppel, S.~Moore, S.~Phillips, D.~Maffitt,
  M.~Pringle, L.~Tarbox, and F.~Prior}.
\newblock \href{https://doi.org/10.1007/s10278-013-9622-7}{The cancer imaging
  archive ({TCIA}): Maintaining and operating a public information repository}.
\newblock \href{https://doi.org/10.1007/s10278-013-9622-7}{{\em Journal of
  Digital Imaging}},
  \href{https://doi.org/10.1007/s10278-013-9622-7}{26(6):1045--1057},
  \href{https://doi.org/10.1007/s10278-013-9622-7}{2013}.
  \href{https://doi.org/10.1007/s10278-013-9622-7}
{doi: {{%
10\hspace{.1pt}\discretionary{.}{%
}{.}\hspace{.4pt}1007\discretionary{/}{%
}{/}s10278\discretionary{%
}{-}{-}013\discretionary{%
}{-}{-}9622\discretionary{%
}{-}{-}7}}}


\bibitem{inProc:JCrypt}
\href{https://doi.org/10.1145/2972206.2972209}{Y.~Dong, A.~Milanova, and
  J.~Dolby}.
\newblock \href{https://doi.org/10.1145/2972206.2972209}{{JCrypt}: Towards
  computation over encrypted data}.
\newblock \href{https://doi.org/10.1145/2972206.2972209}{In {\em Proceedings of
  PPPJ '16}}, \href{https://doi.org/10.1145/2972206.2972209}{2016}.
  \href{https://doi.org/10.1145/2972206.2972209}
{doi: {{%
10\hspace{.1pt}\discretionary{.}{%
}{.}\hspace{.4pt}1145\discretionary{/}{%
}{/}2972206\hspace{.1pt}\discretionary{.}{%
}{.}\hspace{.4pt}2972209}}}


\bibitem{inProc:SecureMR}
\href{https://doi.org/10.1145/3190619.3190638}{Y.~Dong, A.~Milanova, and
  J.~Dolby}.
\newblock \href{https://doi.org/10.1145/3190619.3190638}{{SecureMR}: Secure
  mapreduce computation using homomorphic encryption and program partitioning}.
\newblock \href{https://doi.org/10.1145/3190619.3190638}{In {\em Proceedings of
  HoTSoS '18}}, \href{https://doi.org/10.1145/3190619.3190638}{pp. 841--848},
  \href{https://doi.org/10.1145/3190619.3190638}{2018}.
  \href{https://doi.org/10.1145/3190619.3190638}
{doi: {{%
10\hspace{.1pt}\discretionary{.}{%
}{.}\hspace{.4pt}1145\discretionary{/}{%
}{/}3190619\hspace{.1pt}\discretionary{.}{%
}{.}\hspace{.4pt}3190638}}}


\bibitem{NIST:AES}
\href{https://doi.org/10.6028/NIST.FIPS.197}{M.~J. Dworkin, E.~B. Barker, J.~R.
  Nechvatal, J.~Foti, L.~E. Bassham, E.~Roback, and J.~F. {Dray Jr.}}
\newblock \href{https://doi.org/10.6028/NIST.FIPS.197}{Advanced encryption
  standard ({AES})}.
\newblock \href{https://doi.org/10.6028/NIST.FIPS.197}{Technical Report Federal
  Information Processing Standards Publication 197},
  \href{https://doi.org/10.6028/NIST.FIPS.197}{U.S. Department of Commerce,
  National Institute of Standards and Technology, Gaithersburg, MD},
  \href{https://doi.org/10.6028/NIST.FIPS.197}{2001}.
  \href{https://doi.org/10.6028/NIST.FIPS.197}
{doi: {{%
10\hspace{.1pt}\discretionary{.}{%
}{.}\hspace{.4pt}6028\discretionary{/}{%
}{/}NIST\hspace{.1pt}\discretionary{.}{%
}{.}\hspace{.4pt}FIPS\hspace{.1pt}\discretionary{.}{%
}{.}\hspace{.4pt}197}}}


\bibitem{inproc:Fazio:2018:paillierOptimizingGenerator}
\href{https://doi.org/10.1007/978-3-319-68637-0_23}{N.~Fazio, R.~Gennaro,
  T.~Jafarikhah, and W.~E.~Skeith~III}.
\newblock \href{https://doi.org/10.1007/978-3-319-68637-0_23}{Homomorphic
  secret sharing from paillier encryption}.
\newblock \href{https://doi.org/10.1007/978-3-319-68637-0_23}{In {\em
  Proceedings of 11th Provable Security}},
  \href{https://doi.org/10.1007/978-3-319-68637-0_23}{pp. 381--399},
  \href{https://doi.org/10.1007/978-3-319-68637-0_23}{2017}.
  \href{https://doi.org/10.1007/978-3-319-68637-0_23}
{doi: {{%
10\hspace{.1pt}\discretionary{.}{%
}{.}\hspace{.4pt}1007\discretionary{/}{%
}{/}978\discretionary{%
}{-}{-}3\discretionary{%
}{-}{-}319\discretionary{%
}{-}{-}68637\discretionary{%
}{-}{-}0\_23}}}


\bibitem{phdthesis:homencGentry}
C.~Gentry.
\newblock {\em A fully homomorphic encryption scheme}.
\newblock PhD thesis, Stanford University, 2009.

\bibitem{web:cryptographicKeyLength}
D.~Giry and J.-J. Quisquater.
\newblock Bsi cryptographic key length report (2018) -
  \url{https://www.keylength.com/en/8/}.
\newblock Online.
\newblock Accessed: March 3 2020.

\bibitem{web:GoogleEncryptedBigqueryClientGit}
Google encrypted bigquery client
  \url{https://github.com/google/encrypted-bigquery-client}.
\newblock Online.
\newblock Accessed: April 17 2020.

\bibitem{datasetPresent}
\href{https://www.cg.tuwien.ac.at/research/publications/2006/dataset-present/}{C.~Heinzl}.
\newblock
  \href{https://www.cg.tuwien.ac.at/research/publications/2006/dataset-present/}{Christmas
  present [dataset]
  \url{https://www.cg.tuwien.ac.at/research/publications/2006/dataset-present/}},
  \href{https://www.cg.tuwien.ac.at/research/publications/2006/dataset-present/}{2006}.

\bibitem{article:hoepman2008securityThrough}
\href{https://doi.org/10.1145/1188913.1188921}{J.-H. Hoepman and B.~Jacobs}.
\newblock \href{https://doi.org/10.1145/1188913.1188921}{Increased security
  through open source}.
\newblock \href{https://doi.org/10.1145/1188913.1188921}{{\em Communications of
  the ACM}}, \href{https://doi.org/10.1145/1188913.1188921}{50(1):79--83},
  \href{https://doi.org/10.1145/1188913.1188921}{2007}.
  \href{https://doi.org/10.1145/1188913.1188921}
{doi: {{%
10\hspace{.1pt}\discretionary{.}{%
}{.}\hspace{.4pt}1145\discretionary{/}{%
}{/}1188913\hspace{.1pt}\discretionary{.}{%
}{.}\hspace{.4pt}1188921}}}


\bibitem{phdthesis:Hu2013ImprovingHE}
Y.~Hu.
\newblock {\em Improving the Efficiency of Homomorphic Encryption Schemes}.
\newblock PhD thesis, Worcester Polytechnic Institute, 2013.

\bibitem{web:JavallierGit}
javallier \url{https://github.com/n1analytics/javallier}.
\newblock Online.
\newblock Accessed: April 17 2020.

\bibitem{tech:RFC4301.IPsec}
\href{https://www.rfc-editor.org/rfc/rfc4301.txt}{S.~Kent and K.~Seo}.
\newblock \href{https://www.rfc-editor.org/rfc/rfc4301.txt}{Security
  architecture for the internet protocol}.
\newblock \href{https://www.rfc-editor.org/rfc/rfc4301.txt}{RFC 4301},
  \href{https://www.rfc-editor.org/rfc/rfc4301.txt}{2005}.
\newblock
  \href{https://www.rfc-editor.org/rfc/rfc4301.txt}{\url{https://www.rfc-editor.org/rfc/rfc4301.txt}}.

\bibitem{article:Kerckhoffs83}
\href{http://www.petitcolas.net/fabien/kerckhoffs/}{A.~Kerckhoffs}.
\newblock \href{http://www.petitcolas.net/fabien/kerckhoffs/}{{La cryptographie
  militaire}}.
\newblock \href{http://www.petitcolas.net/fabien/kerckhoffs/}{{\em Journal des
  sciences militaires}},
  \href{http://www.petitcolas.net/fabien/kerckhoffs/}{IX:5--38},
  \href{http://www.petitcolas.net/fabien/kerckhoffs/}{1883}.

\bibitem{book:KnuthArtV2}
D.~E. Knuth.
\newblock {\em The art of computer programming}, vol.~2, chap. 4.5.2 The
  Greatest Common Divisor, p. 325.
\newblock Addison-Wesley, Reading, MA, US, 2 ed., 1981.

\bibitem{proc:Krueger:2003:ATGV}
\href{https://doi.org/10.1109/VISUAL.2003.1250384}{J.~Kr{\"u}ger and
  R.~Westermann}.
\newblock \href{https://doi.org/10.1109/VISUAL.2003.1250384}{Acceleration
  techniques for {GPU}-based {Volume} {Rendering}}.
\newblock \href{https://doi.org/10.1109/VISUAL.2003.1250384}{In {\em
  Proceedings of VIS'03}},
  \href{https://doi.org/10.1109/VISUAL.2003.1250384}{pp. 287--292},
  \href{https://doi.org/10.1109/VISUAL.2003.1250384}{2003}.
  \href{https://doi.org/10.1109/VISUAL.2003.1250384}
{doi: {{%
10\hspace{.1pt}\discretionary{.}{%
}{.}\hspace{.4pt}1109\discretionary{/}{%
}{/}VISUAL\hspace{.1pt}\discretionary{.}{%
}{.}\hspace{.4pt}2003\hspace{.1pt}\discretionary{.}{%
}{.}\hspace{.4pt}1250384}}}


\bibitem{datasetLungCancer}
\href{https://doi.org/10.7937/TCIA.2020.NNC2-0461}{P.~Li, S.~Wang, T.~Li,
  J.~Lu, Y.~HuangFu, and D.~Wang}.
\newblock \href{https://doi.org/10.7937/TCIA.2020.NNC2-0461}{A large-scale {CT}
  and {PET/CT} dataset for lung cancer diagnosis [dataset]. {The Cancer Imaging
  Archive}}, \href{https://doi.org/10.7937/TCIA.2020.NNC2-0461}{2020}.
  \href{https://doi.org/10.7937/TCIA.2020.NNC2-0461}
{doi: {{%
10\hspace{.1pt}\discretionary{.}{%
}{.}\hspace{.4pt}7937\discretionary{/}{%
}{/}TCIA\hspace{.1pt}\discretionary{.}{%
}{.}\hspace{.4pt}2020\hspace{.1pt}\discretionary{.}{%
}{.}\hspace{.4pt}NNC2\discretionary{%
}{-}{-}0461}}}


\bibitem{article:MartinsFheSurvey2018}
\href{https://doi.org/10.1145/3124441}{P.~Martins, L.~Sousa, and A.~Mariano}.
\newblock \href{https://doi.org/10.1145/3124441}{A survey on fully homomorphic
  encryption: An engineering perspective}.
\newblock \href{https://doi.org/10.1145/3124441}{{\em ACM Computing Surveys}},
  \href{https://doi.org/10.1145/3124441}{50(6):83:1--83:33},
  \href{https://doi.org/10.1145/3124441}{2017}.
  \href{https://doi.org/10.1145/3124441}
{doi: {{%
10\hspace{.1pt}\discretionary{.}{%
}{.}\hspace{.4pt}1145\discretionary{/}{%
}{/}3124441}}}


\bibitem{inproc:3DCrypt}
\href{https://doi.org/10.5220/0005966302830291}{M.~Mohanty, M.~R. Asghar, and
  G.~Russello}.
\newblock \href{https://doi.org/10.5220/0005966302830291}{3dcrypt:
  Privacy-preserving pre-classification volume ray-casting of 3d images in the
  cloud}.
\newblock \href{https://doi.org/10.5220/0005966302830291}{In {\em Proceedings
  of 13th E-Business and Telecommunications}},
  \href{https://doi.org/10.5220/0005966302830291}{pp. 283--291},
  \href{https://doi.org/10.5220/0005966302830291}{2016}.
  \href{https://doi.org/10.5220/0005966302830291}
{doi: {{%
10\hspace{.1pt}\discretionary{.}{%
}{.}\hspace{.4pt}5220\discretionary{/}{%
}{/}0005966302830291}}}


\bibitem{tech:mortonOder}
G.~Morton.
\newblock A computer oriented geodetic data base and a new technique in file
  sequencing.
\newblock Technical report, IBM Co. Ltd., 150 Laurier Avenue, West, Ottawa 4,
  Ontario, Canada, 1966.

\bibitem{pdf:Paillier}
\href{https://doi.org/10.1007/3-540-48910-X_16}{P.~Paillier}.
\newblock \href{https://doi.org/10.1007/3-540-48910-X_16}{Public-key
  cryptosystems based on composite degree residuosity classes}.
\newblock \href{https://doi.org/10.1007/3-540-48910-X_16}{In {\em Procceedings
  of {EUROCRYPT} '99}}, \href{https://doi.org/10.1007/3-540-48910-X_16}{pp.
  223--238}, \href{https://doi.org/10.1007/3-540-48910-X_16}{1999}.
  \href{https://doi.org/10.1007/3-540-48910-X_16}
{doi: {{%
10\hspace{.1pt}\discretionary{.}{%
}{.}\hspace{.4pt}1007\discretionary{/}{%
}{/}3\discretionary{%
}{-}{-}540\discretionary{%
}{-}{-}48910\discretionary{%
}{-}{-}X\_16}}}


\bibitem{honestButCurious}
\href{https://www.cs.ox.ac.uk/people/andrew.paverd/casper/casper-privacy-report.pdf}{A.~Paverd,
  A.~Martin, and I.~Brown}.
\newblock
  \href{https://www.cs.ox.ac.uk/people/andrew.paverd/casper/casper-privacy-report.pdf}{Modelling
  and automatically analysing privacy properties for honest-but-curious
  adversaries}.
\newblock
  \href{https://www.cs.ox.ac.uk/people/andrew.paverd/casper/casper-privacy-report.pdf}{Technical
  report},
  \href{https://www.cs.ox.ac.uk/people/andrew.paverd/casper/casper-privacy-report.pdf}{2014}.
\newblock
  \href{https://www.cs.ox.ac.uk/people/andrew.paverd/casper/casper-privacy-report.pdf}{Online.
  \url{https://www.cs.ox.ac.uk/people/andrew.paverd/casper/casper-privacy-report.pdf}
  Accessed: July 30 2019}.

\bibitem{web:ChadPerrinSecurity101}
C.~Perrin.
\newblock Security 101, remedial edition: Obscurity is not security -
  \url{https://www.techrepublic.com/blog/it-security/security-101-remedial-edition-obscurity-is-not-security/}.
\newblock Online.
\newblock Accessed: April 17 2020.

\bibitem{Porter1984}
\href{https://doi.org/10.1145/964965.808606}{T.~Porter and T.~Duff}.
\newblock \href{https://doi.org/10.1145/964965.808606}{Compositing digital
  images}.
\newblock \href{https://doi.org/10.1145/964965.808606}{In {\em Proceedings of
  SIGGRAPH '84}}, \href{https://doi.org/10.1145/964965.808606}{pp.
  253–--259}, \href{https://doi.org/10.1145/964965.808606}{1984}.
  \href{https://doi.org/10.1145/964965.808606}
{doi: {{%
10\hspace{.1pt}\discretionary{.}{%
}{.}\hspace{.4pt}1145\discretionary{/}{%
}{/}964965\hspace{.1pt}\discretionary{.}{%
}{.}\hspace{.4pt}808606}}}


\bibitem{web:PythonPaillierGit}
python-paillier \url{https://github.com/n1analytics/python-paillier}.
\newblock Online.
\newblock Accessed: April 17 2020.

\bibitem{tech:RFC8446.TLS}
\href{https://www.rfc-editor.org/rfc/rfc8446.txt}{E.~Rescorla}.
\newblock \href{https://www.rfc-editor.org/rfc/rfc8446.txt}{The transport layer
  security ({TLS}) protocol version 1.3}.
\newblock \href{https://www.rfc-editor.org/rfc/rfc8446.txt}{RFC 8446},
  \href{https://www.rfc-editor.org/rfc/rfc8446.txt}{2018}.
\newblock
  \href{https://www.rfc-editor.org/rfc/rfc8446.txt}{\url{https://www.rfc-editor.org/rfc/rfc8446.txt}}.

\bibitem{pdf:FirstHE.Rivest1978}
R.~L. Rivest, L.~Adleman, and M.~L. Dertouzos.
\newblock On data banks and privacy homomorphisms.
\newblock {\em Foundations of Secure Computation, Academia Press},
  4(11):169--179, 1978.

\bibitem{article:rsa}
\href{https://doi.org/10.1145/359340.359342}{R.~L. Rivest, A.~Shamir, and
  L.~Adleman}.
\newblock \href{https://doi.org/10.1145/359340.359342}{A method for obtaining
  digital signatures and public-key cryptosystems}.
\newblock \href{https://doi.org/10.1145/359340.359342}{{\em Communications of
  the ACM}}, \href{https://doi.org/10.1145/359340.359342}{21(2):120--126},
  \href{https://doi.org/10.1145/359340.359342}{1978}.
  \href{https://doi.org/10.1145/359340.359342}
{doi: {{%
10\hspace{.1pt}\discretionary{.}{%
}{.}\hspace{.4pt}1145\discretionary{/}{%
}{/}359340\hspace{.1pt}\discretionary{.}{%
}{.}\hspace{.4pt}359342}}}


\bibitem{NIST:800_123}
\href{https://doi.org/10.6028/NIST.SP.800-123}{K.~Scarfone, W.~Jansen, and
  M.~Tracy}.
\newblock \href{https://doi.org/10.6028/NIST.SP.800-123}{Guide to general
  server security}.
\newblock \href{https://doi.org/10.6028/NIST.SP.800-123}{Technical Report NIST
  Special Publication 800-123},
  \href{https://doi.org/10.6028/NIST.SP.800-123}{U.S. Department of Commerce,
  National Institute of Standards and Technology, Gaithersburg, MD},
  \href{https://doi.org/10.6028/NIST.SP.800-123}{2018}.
  \href{https://doi.org/10.6028/NIST.SP.800-123}
{doi: {{%
10\hspace{.1pt}\discretionary{.}{%
}{.}\hspace{.4pt}6028\discretionary{/}{%
}{/}NIST\hspace{.1pt}\discretionary{.}{%
}{.}\hspace{.4pt}SP\hspace{.1pt}\discretionary{.}{%
}{.}\hspace{.4pt}800\discretionary{%
}{-}{-}123}}}


\bibitem{web:BruceSchneierTheInsecurity2}
B.~Schneier.
\newblock The insecurity of secret it systems -
  \url{https://www.schneier.com/blog/archives/2014/02/the_insecurity_2.html}.
\newblock Online.
\newblock Accessed: April 17 2020.

\bibitem{inProc:Crypsis}
\href{https://www.usenix.org/conference/hotcloud14/workshop-program/presentation/stephen}{J.~J.
  Stephen, S.~Savvides, R.~Seidel, and P.~Eugster}.
\newblock
  \href{https://www.usenix.org/conference/hotcloud14/workshop-program/presentation/stephen}{Practical
  confidentiality preserving big data analysis}.
\newblock
  \href{https://www.usenix.org/conference/hotcloud14/workshop-program/presentation/stephen}{In
  {\em Proceedings of HotCloud'14}},
  \href{https://www.usenix.org/conference/hotcloud14/workshop-program/presentation/stephen}{2014}.

\bibitem{inProc:GPUMapReduce}
\href{https://doi.org/10.1145/1851476.1851597}{J.~A. Stuart, C.-K. Chen, K.-L.
  Ma, and J.~D. Owens}.
\newblock \href{https://doi.org/10.1145/1851476.1851597}{Multi-{GPU} volume
  rendering using {MapReduce}}.
\newblock \href{https://doi.org/10.1145/1851476.1851597}{In {\em Proceedings of
  ACM HPDC '10}}, \href{https://doi.org/10.1145/1851476.1851597}{pp. 841--848},
  \href{https://doi.org/10.1145/1851476.1851597}{2010}.
  \href{https://doi.org/10.1145/1851476.1851597}
{doi: {{%
10\hspace{.1pt}\discretionary{.}{%
}{.}\hspace{.4pt}1145\discretionary{/}{%
}{/}1851476\hspace{.1pt}\discretionary{.}{%
}{.}\hspace{.4pt}1851597}}}


\bibitem{pdf:ObliviousLookupTables}
\href{https://doi.org/10.1515/tmmp-2016-0039}{M.~S. Wamser, S.~Rass, and
  P.~Schartner}.
\newblock \href{https://doi.org/10.1515/tmmp-2016-0039}{Oblivious
  lookup-tables}.
\newblock \href{https://doi.org/10.1515/tmmp-2016-0039}{{\em Tatra Mountains
  Mathematical Publications}},
  \href{https://doi.org/10.1515/tmmp-2016-0039}{67(1):191--203},
  \href{https://doi.org/10.1515/tmmp-2016-0039}{2016}.
  \href{https://doi.org/10.1515/tmmp-2016-0039}
{doi: {{%
10\hspace{.1pt}\discretionary{.}{%
}{.}\hspace{.4pt}1515\discretionary{/}{%
}{/}tmmp\discretionary{%
}{-}{-}2016\discretionary{%
}{-}{-}0039}}}


\bibitem{inbook:homoEncApps:2014:paillier}
\href{https://doi.org/10.1007/978-3-319-12229-8}{X.~Yi, R.~Paulet, and
  E.~Bertino}.
\newblock \href{https://doi.org/10.1007/978-3-319-12229-8}{{\em Homomorphic
  Encryption and Applications}},
  \href{https://doi.org/10.1007/978-3-319-12229-8}{chap. 2 Homomorphic
  Encryption, p.~41}.
\newblock \href{https://doi.org/10.1007/978-3-319-12229-8}{Springer},
  \href{https://doi.org/10.1007/978-3-319-12229-8}{2014}.
  \href{https://doi.org/10.1007/978-3-319-12229-8}
{doi: {{%
10\hspace{.1pt}\discretionary{.}{%
}{.}\hspace{.4pt}1007\discretionary{/}{%
}{/}978\discretionary{%
}{-}{-}3\discretionary{%
}{-}{-}319\discretionary{%
}{-}{-}12229\discretionary{%
}{-}{-}8}}}


\bibitem{pdf:CryptoImg-IEEE}
\href{https://doi.org/10.1109/CNS.2016.7860550}{M.~T.~I. Ziad, A.~Alanwar,
  M.~Alzantot, and M.~Srivastava}.
\newblock \href{https://doi.org/10.1109/CNS.2016.7860550}{Cryptoimg: Privacy
  preserving processing over encrypted images}.
\newblock \href{https://doi.org/10.1109/CNS.2016.7860550}{In {\em Procceedings
  of Communications and Network Security}},
  \href{https://doi.org/10.1109/CNS.2016.7860550}{pp. 570--575},
  \href{https://doi.org/10.1109/CNS.2016.7860550}{2016}.
  \href{https://doi.org/10.1109/CNS.2016.7860550}
{doi: {{%
10\hspace{.1pt}\discretionary{.}{%
}{.}\hspace{.4pt}1109\discretionary{/}{%
}{/}CNS\hspace{.1pt}\discretionary{.}{%
}{.}\hspace{.4pt}2016\hspace{.1pt}\discretionary{.}{%
}{.}\hspace{.4pt}7860550}}}


\end{thebibliography}

\appendix

\clearpage
\section*{Supplementary Material to the Submission \enquote{Homomorphic-Encrypted Volume Rendering}}


\section{Background on Homomorphic Encryption}
\label{sec:HE}
{\em Homomorphic Encryption} schemes allow computations on encrypted data such that the decrypted results are equal to the result of a mathematical or logical operation applied to the corresponding plaintext data.
Therefore, calculations on encrypted data can be performed without decrypting the data first.
This property makes it possible to outsource not only the storage of encrypted data, but also the computation on sensitive data to untrusted third parties (e.g., the cloud).
The computation on cloud servers has two major advantages.
If only the result and not the whole dataset needs to be transferred to the client, a lower network bandwidth is required. Furthermore, the client can be a thin client like a tablet without much computing and storage resources because the computational expensive rendering is done on the server.
Homomorphic encryption schemes are classified into three categories:
\begin{itemize}
	\itemsep0em
	\item {\em partially homomorphic encryption (PHE)}: are homomorphic with regard to only one type of operation (addition or multiplication)
	\item {\em somewhat homomorphic encryption (SHE)}: Can perform more general calculations than PHE, but only a limited number of them.
	\item {\em fully homomorphic encryption (FHE)}: Can perform any computation on encrypted data.
\end{itemize}
Rivest et al. \cite{pdf:FirstHE.Rivest1978} invented the idea of Homomorphic Encryption in 1978. They showed the demand of a secure Homomorphic Encryption scheme that supports a large set of operations.
However, it took more than 30 years until the first proposal of such a FHE was made by C. Gentry in 2009 \cite{phdthesis:homencGentry}.
While the first FHE schemes were just concepts, due to an ongoing development and optimization process, they are currently at least efficient enough for functional implementations.

However, FHE schemes are still not practical for real applications, because the storage and computational costs are too high \cite{phdthesis:Hu2013ImprovingHE, article:MartinsFheSurvey2018, article:AbbasHeSurvey2018}.
Therefore, in our approach, we propose to take advantage of the Paillier PHE scheme, whose homomorphic properties that are relevant for our encrypted volume rendering will be introduced next.

\subsection{The Paillier Cryptosystem}
\label{sec:PaillierCryptosystem}
Paillier's cryptosystem \cite{pdf:Paillier} is an additive PHE scheme.
The important property of this scheme is that a multiplication of two encrypted numbers is equivalent to an addition in the plaintext domain.
This means that it is possible to calculate the sum of plaintext numbers that have been encrypted by multiplying the encrypted numbers.
This relation is stated in \autoref{equ:paillierAddition}.
For \autoref{equ:paillierAddition} and  \autoref{equ:paillierMultiplication}, we adopt the  notation by Ziad et al. \cite{pdf:CryptoImg-IEEE}.
The $\oplus$ symbol is used for the operation on encrypted numbers that is equivalent to an addition on plaintext numbers.
The encrypted version of the value $m$ is denoted as $\llbracket m \rrbracket$ and $\textrm{Dec}(\llbracket m \rrbracket) = m$ means decrypting $\llbracket m \rrbracket$ to $m$ by the decryption function of Paillier's cryptosystem (see: \autoref{alg:PaillierDecrypt}).
Paillier works on finite fields and thus performs a modulo (mod $N^2$) operation after each multiplication ensures to stay in the field and does not change the decrypted result because the corresponding plaintext numbers need to be less than the modulus $N$ for a correct decryption anyway.
However, the modulo operation makes further calculations more efficient because it prevents unnecessary large numbers ($n$ multiplications will blow up the number length by about $n$ times).
%
\begin{align}
	\begin{split}
		\textrm{Dec}(\llbracket m_1 \rrbracket \oplus \llbracket m_2 \rrbracket) &= \textrm{Dec}((\llbracket m_1 \rrbracket \times \llbracket m_2 \rrbracket) \mod N^2)\\
		&= (m_1 + m_2) \mod N
	\end{split}
	\label{equ:paillierAddition}
\end{align}
Since it is possible to add encrypted numbers to each other, in a \texttt{for-loop}, an encrypted number can be added to itself $d$ times to simulate a multiplication with $d$.
However, note that the value $d$ is in plaintext.
Since addition is performed by doing multiplication on the encrypted numbers, we can, instead of a \texttt{for-loop}, take the encrypted value to the power of $d$ to get this result, which can be implemented more efficiently than a \texttt{for-loop}.
Furthermore, it has the advantage that it also works for $d < 0$.
The case $d = -1$ is of special interest, because this makes subtraction of two encrypted numbers possible ($\textrm{Dec}(\llbracket m_1 - m_2 \rrbracket) = \textrm{Dec}( \llbracket m_1 \rrbracket \times \llbracket m_2 \rrbracket^{-1} \: \mod \: N^2)$).
The symbol $\otimes$ is used for such an operation on one plaintext number $d$ and one encrypted number $\llbracket m_1 \rrbracket$.
\autoref{equ:paillierMultiplication} shows how to calculate this multiplication with one encrypted number.
%
\begin{align}
\begin{split}
	\textrm{Dec}(\llbracket m_1 \rrbracket \otimes d) &= \textrm{Dec}(\llbracket m_1 \rrbracket^d \mod N^2)\\
	&= (m_1 \times d) \mod N
\end{split}
\label{equ:paillierMultiplication}
\end{align}
While the Paillier PHE supports an efficient method to multiply an encrypted and a plaintext number, it does not support the multiplication of two encrypted numbers and is, therefore, not a fully homomorphic encryption scheme.


\begin{algorithm}[tb]
\DontPrintSemicolon

\SetKwInOut{Param}{Parameters}
\SetKw{TypeInt}{integer}
\SetKwProg{Proc}{procedure}{}{}

\SetKwFunction{FnCreate}{create}
\SetKwFunction{FnBitLength}{bitLength}
\SetKwData{PublicKey}{PublicKey}
\SetKwData{SecureKey}{SecureKey}

\BlankLine
\Param{Length of the modulus $N$ in bit ($b$).}

\KwResult{The secure key ($p$, $q$) and the public key ($N$, $g$).}

\BlankLine

\Proc{\FnCreate{$b$}}{
	\Repeat{\FnBitLength{$N$} $=$ $b$}{
		$p = $ random prime number with a length of $\frac{b}{2}$ bits \;
		$q = $ random prime number $\neq p$ with a length of $\frac{b}{2}$ bits \;

		$N = p * q$ \;
	}
	$g = N + 1$ \;

	\Return $(p, q)$ and $(N, g)$ \;
}
\caption{Paillier Create Keys}
\label{alg:PaillierCreateKeys}
\end{algorithm}


\begin{algorithm}[tb]
\DontPrintSemicolon

\SetKwInOut{Param}{Parameters}
\SetKw{TypeInt}{integer}
\SetKwProg{Proc}{procedure}{}{}

\SetKwFunction{FnEncrypt}{encrypt}
\SetKwFunction{FnBitLength}{bitLength}
\SetKwData{PublicKey}{PublicKey}

\BlankLine

\Param{The plaintext message $m$ ($ < N$) that should be encrypted with the public key ($N$, $g$).}

\KwResult{Ciphertext $c$}
\BlankLine

\Proc{\FnEncrypt{$m$, $N$, $g$}}{
	$r = $ random number that is smaller than $N$ ($r < N$) \;
	$c = g^m \cdot r^N \mod N^2$ \;
	\Return $c$ \;
}
\caption{Paillier Encrypt}
\label{alg:PaillierEncrypt}
\end{algorithm}


\begin{algorithm}[tb]
\DontPrintSemicolon

\SetKwInOut{Param}{Parameters}
\SetKw{TypeInt}{integer}
\SetKwProg{Proc}{procedure}{}{}

\SetKwFunction{FnDecrypt}{decrypt}
\SetKwFunction{FnBitLength}{bitLength}
\SetKwData{PublicKey}{PublicKey}
\SetKwData{SecureKey}{SecureKey}

\BlankLine

\Param{The ciphertext $c$ that should be decrypted with the secure key ($p$, $q$) and the public key ($N$, $g$).}

\KwResult{Plaintext $m$}
\BlankLine

\Proc{\FnDecrypt{$c$, $p$, $q$, $N$, $g$}}{
	\tcp{ L- and H-function as defined by Paillier\cite{pdf:Paillier} at section 7}
	\Proc {$L(x, y)$} {
		\Return $\frac{x-1}{y}$ \;
	}
	\BlankLine
	\Proc {$H(x)$} {
		\Return $ L(g^{x-1} \textrm{ mod } x^2, x)^{-1} \mod x$ \;
	}
	\BlankLine

	\Proc(\tcp*[f]{Chinese Remainder Theorem}) {$C(m_p, m_q)$} {
		\Return $m_p + (((m_q - m_p) \cdot (p^{-1} \textrm{ mod } q ) \textrm{ mod } q) \cdot p)$ \;
	}
	\BlankLine

	$h_p = H(p)$ \tcp*{can be pre computed}
	$h_q = H(q)$ \tcp*{can be pre computed}
	$m_p = L(c^{p-1} \textrm{ mod } p^2, p) \cdot h_p \textrm{ mod } p$ \;
	$m_q = L(c^{q-1} \textrm{ mod } q^2, q) \cdot h_q \textrm{ mod } q$ \;
	$m = C(m_p, m_q)$ \;

	\Return $m$ \;
}
\caption{Paillier Decrypt}
\label{alg:PaillierDecrypt}
\end{algorithm}

The Paillier HE is a probabilistic asymmetric encryption scheme like the well-known RSA scheme \cite{article:rsa}.
That means that an encryption can be performed by using the public key, which is derived from the secure (or private) key.
However, for the task of decryption, the secure key is required, which cannot (efficiently) be calculated from the public key because this would require factoring a product of two large prime numbers.
It is essential for the security of Paillier's cryptosystem (and also for RSA) that there is no known fast method for integer factorization of a product of two large prime numbers.
The \autoref{alg:PaillierCreateKeys} shows an example of a key generation function for the Paillier cryptosystem, which sets the generator $g$ always to $N+1$, as this allows a more efficient encryption function (see: \cite{inproc:Fazio:2018:paillierOptimizingGenerator}).
Furthermore, the algorithm clearly shows that the secure key, which needs to be kept secret, contains the two large prime numbers $p$ and $q$ (e.g., 1024 bit long) and the public key contains the product $N$ (e.g., 2048 bit long) of these two prime numbers.

\autoref{alg:PaillierEncrypt} and \autoref{alg:PaillierDecrypt} show the pseudocode for the encrypt and decrypt routines for the Paillier HE.
The encryption algorithm also contains the obfuscating of an encrypted number with a random number $r$, which qualifies the Paillier HE as a probabilistic encryption scheme.
This means that a specific plaintext message $m$ can be represented by many possible ciphertexts $ \llbracket m \rrbracket_1,  \llbracket m \rrbracket_2,  \llbracket m \rrbracket_3 \dots,  \llbracket m \rrbracket_r$.
The decryption with the right secure key will return the original message $m$ for all the possible ciphertext representations.
While this is not required for correct homomorphic calculations with Paillier (imagine $r = 1$), it is important for the semantic security against chosen-plaintext attacks (IND-CPA) that Paillier's cryptosystem provides \cite{pdf:Paillier, inbook:homoEncApps:2014:paillier}.
Without this obfuscating, it would be possible to decrypt datasets without knowing the secure key because an attacker would only need to encrypt all possible plaintext values with the public key,
store the plaintext and the corresponding ciphertext values in pairs, and compare the values of the encrypted dataset with the self-encrypted values for which the correct decryption is known.
For datasets with a limited number of possible values like the voxel values of a volume, which usually contains no more than $2^{10} = 1024$ different values, this would be a trivial task.

\end{document}